\documentclass[journal]{IEEEtran}
\usepackage[T1]{fontenc}
\usepackage[latin9]{inputenc}
\usepackage{amsthm}
\usepackage{amsmath} 
\usepackage{graphicx} 
\usepackage{color} 
\usepackage{cite}
\usepackage{algpseudocode}
\usepackage{algorithm}
\usepackage{amssymb}
\usepackage{subfigure}
\usepackage{esint}
\usepackage{algpseudocode}
\usepackage{epstopdf}
\usepackage{multirow}
\usepackage{makecell}
\usepackage{float}
\usepackage{lipsum}
\usepackage{balance}
\usepackage{bbm}
\usepackage{tikz}
\usepackage{makecell} 

\usetikzlibrary{decorations,decorations.markings}

\tikzset{
  strike through/.style={
    postaction=decorate,
    decoration={
      markings,
      mark=at position 0.5 with {
        \draw[-] (0pt,-3pt) -- (0pt, 3pt);
      }
    }
  }
}

\usetikzlibrary{arrows}
\usetikzlibrary{automata} 
\usetikzlibrary{positioning} 
\tikzstyle{line} = [draw,thick,-latex]
\tikzstyle{transition} = [font=\small]

\definecolor{green}{rgb}{0.13, 0.55, 0.13}
\definecolor{brown}{rgb}{0.6, 0.2, 0.0}

\newtheorem{definition}{Definition}
\newtheorem{remark}{Remark}

\newtheorem{example}{Example}

\theoremstyle{plain}
\theoremstyle{plain}
\newtheorem{theorem}{Theorem}
\newtheorem{lemma}{Lemma}
\newtheorem{notation}{Notation}

\newtheorem{proposition}{Proposition}

\newcommand{\be}{\begin{eqnarray}}
\newcommand{\ee}{\end{eqnarray}}
\newcommand{\comment}[1]{}

\IEEEoverridecommandlockouts

\allowdisplaybreaks

\begin{document}

\title{Protecting the Future of Information: LOCO Coding With Error Detection for DNA Data Storage
}

\author{
   \IEEEauthorblockN{Canberk \.{I}rima\u{g}z{\i}, Yusuf Uslan, and Ahmed Hareedy, \IEEEmembership{Member, IEEE}} \vspace{-1.0em}
   
   \thanks{This work was supported in part by the T\"{U}B\.ITAK 2232-B International Fellowship for Early Stage Researchers. 
\par
Canberk \.{I}rima\u{g}z{\i} is with the Institute of Applied Mathematics, Middle East Technical University (METU), 06800 Ankara, Turkey (e-mail: canberk.irimagzi@metu.edu.tr).
\par
Yusuf Uslan and Ahmed Hareedy are with the Department of Electrical and Electronics Engineering, Middle East Technical University, 06800 Ankara, Turkey (e-mail: yusuf.uslan@metu.edu.tr; ahareedy@metu.edu.tr).}
}
\maketitle

%%%%%%%%%%%%%%%%%%%%%%%%%%%%%%%%%%%%%
\begin{abstract}
From the information-theoretic perspective, DNA strands serve as a storage medium for $4$-ary data over the alphabet $\{A,T,G,C\}$. DNA data storage promises formidable information density, long-term durability, and ease of replicability. However, information in this intriguing storage technology might be corrupted because of error-prone data sequences as well as insertion, deletion, and substitution errors. Experiments have revealed that DNA sequences with long homopolymers and/or with low $GC$-content are notably more subject to errors upon storage. In order to address this biochemical challenge, constrained codes are proposed for usage in DNA data storage systems, and they are studied in the literature accordingly.

This paper investigates the utilization of the recently-introduced method for designing lexicographically-ordered constrained (LOCO) codes in DNA data storage to improve performance. LOCO codes offer capacity-achievability, low complexity, and ease of reconfigurability. This paper introduces novel constrained codes, namely DNA LOCO (D-LOCO) codes, over the alphabet $\{A,T,G,C\}$ with limited runs of identical symbols. Due to their ordered structure, these codes come with an encoding-decoding rule we derive, which provides simple and affordable encoding-decoding algorithms. In terms of storage overhead, the proposed encoding-decoding algorithms outperform those in the existing literature. Our algorithms are based on small-size adders, and therefore they are readily reconfigurable. D-LOCO codes are intrinsically balanced, which allows us to achieve balanced $AT$- and $GC$-content over the entire DNA strand with minimal rate penalty. Moreover, we propose four schemes to bridge consecutive codewords, three of which guarantee single substitution error detection per codeword. We examine the probability of undetecting errors over a presumed symmetric DNA storage channel subject to substitution errors only. We also show that D-LOCO codes are capacity-achieving and that they offer remarkably high rates even at moderate lengths.

\end{abstract}

\begin{IEEEkeywords}
Constrained codes, low-complexity algorithms, reconfigurable coding, LOCO codes, homopolymer run, balancing, error-detection, DNA data storage.
\end{IEEEkeywords}

%%%%%%%%%%%%%%%%%%%%%%%%%%%%%%%%%%%%%
\section{Introduction}\label{sec_intro}

In the current information age, DNA data storage is the next-generation technology offered to accommodate the needs of storing mass cold data \cite{ross_etal}. Due to its remarkable advantage in terms of information density, durability, and ease of replicability over current commercial storage technologies, a pool of synthetic DNA is proposed as a potential medium to store data for archival purposes. Coding and data processing are essential for such emerging technology in order to prevent and correct errors resulting from biochemical effects. To achieve high reliability of DNA strands for a long period of time, sequences with (i) limited runs of identical symbols and (ii) balanced percentage of $A$-$T$ and $G$-$C$ nucleotides are required to be generated for synthesis, i.e., for writing \cite{ross_etal, schwartz_etal}.
\par
Constrained codes are a class of nonlinear codes that~eliminate a chosen set of forbidden patterns from codewords. The use of constrained codes forbidding error-prone patterns in accordance with channel requirements improves system performance, and thus they have a profusion of applications. Historically, in his seminal work of 1948 \cite{shan_const}, Shannon showed how to represent an infinite sequence that forbids certain data patterns through a finite-state transition diagram (FSTD) and defined the capacity, i.e., the highest achievable rate. {In 1970, Tang and Bahl \cite{tang_bahl} as well as Franaszek \cite{franaszek} introduced an important class of constrained codes called run-length-limited (RLL) codes. A $(d,k)$ sequence, or an RLL sequence, is a binary sequence with the constraint that at least $d$ and at most $k$ zeros must separate consecutive ones (see Remark~\ref{remark_RLL} in the Appendix for the relation between the binary version of D-LOCO codes and $(0,k)$ RLL codes).} Since the works \cite{tang_bahl} and \cite{franaszek}, advances in graph theory, numerical matrix theory, and symbolic dynamics remarkably benefited the analysis and design of constrained codes, especially those that are based on finite-state machines \cite{ach_fsm}.
\par
The data storage evolution has always been associated with advances in coding and signal processing. Here, we focus on the value constrained coding has been bringing to various data storage systems. In one-dimensional magnetic recording (MR) systems, constrained codes enabled remarkable density increases in early generations of MR devices that are based on peak detection \cite{siegel_mr}, and they are still used to improve performance in MR systems adopting sequence detection \cite{siegel_const}. By forbidding detrimental two-dimensional isolation patterns, constrained codes mitigate interference in modern two-dimensional MR systems \cite{sharov_TCon, kato_TCon, bd_tdmr}. As for solid-state storage systems, constrained codes are used to protect Flash memories from the effect of inter-cell interference resulting from charge propagation \cite{lee_ici, veeresh_mlc, ahh_mdl, ahh_ps}. Constrained codes find application in other storage systems such as optical recording devices \cite{immink_surv} and DNA data storage systems \cite{li_zhou_zou, heckel_etal}, which are the topic of this paper. The spectral analysis of constrained codes characterizes desirable properties, such as balancing, in the system of interest \cite{immink_surv, ahh_spectra}.
\par
{
The design of constrained codes by adopting lexicographic indexing, also called enumerative coding, has started with the work of Tang and Bahl \cite{tang_bahl} as well as that of Cover \cite{cover_lex}, and it has been intermittently revisited in history \cite{blake_enum, gu_lex, braun_immink}. Recently, Hareedy, Dabak, and Calderbank \cite{ahh_general}, following the works \cite{ahh_asym, ahh_loco} by Hareedy and Calderbank, introduced a novel technique in the field of constrained coding based on lexicographic indexing. They presented a general method for designing what are now called lexicographically-ordered constrained (LOCO) codes in a systematic manner for any finite set of forbidden patterns. They provided simple formulae, namely encoding-decoding rules, for the index of a codeword in terms of the codeword symbols and the cardinalities of LOCO codes having smaller lengths. In this paper, we follow this step-by-step outline to design constrained codes for DNA data storage with affordable encoding-decoding algorithms in accordance with channel requirements.
} 
\par
LOCO codes are constrained codes having codebooks equipped with lexicographic ordering, and thus they naturally come with unique encoding-decoding algorithms that are based on a rule executed by small-size adders. Therefore, LOCO codes do not require any look-up tables. D-LOCO codes, which are LOCO codes defined over the alphabet $\{A,T,G,C\}$ that eliminate long runs of identical symbols and achieve $AT$- and $GC$-content balance, are proposed here as capacity-achieving LOCO codes for DNA storage systems. Note that the ``D'' in ``D-LOCO'' is simply for ``DNA''.

\subsection{Some Related Works}

Immink and Cai \cite{immink_cai}, \cite{immink_cai_2} studied constrained coding schemes that address $GC$-content balance and homopolymer run-length constraints based on look-up tables. Song et al. \cite{song_cai_etal} provided a high-rate coding scheme that produces DNA sequences with homopolymers of length at most $3$ and whose $GC$-content is probabilistically proved (and empirically verified) to be close to $0.5$ as a fraction. Moreover, they provided a low-complexity encoding-decoding algorithm that is also based on look-up tables. Improving the result in \cite{song_cai_etal}, Wang et al. \cite{wang_etal} offered an efficient coding scheme, a product of which is a constrained code with normalized rate $0.9585$ that eliminates homopolymers of length at least $4$ and that has a guaranteed $40\%-60\%$ $GC$-content based on look-up tables.\footnote{We adopt normalized rates throughout this paper, where the actual DNA code rates are divided by $2$, since they immediately give the fraction of non-redundant information.}
\par
Using a sequence replacement technique, Nguyen et al. \cite{nguyen_etal} proposed low-complexity encoders that convert binary sequences into $4$-ary sequences of limited run-length $\ell$ and $GC$-content in $[0.5 - \epsilon, 0.5+\epsilon]$, where $\epsilon$ is adaptable, which enable correcting a single insertion, single deletion, or single substitution error. For example, for $\ell=4$ and $40\%-60\%$ $GC$-content, such DNA sequences of length $n=100$ symbols offer normalized code rate $0.91$ with no error-correction. {In \cite{Nguyen_siegel}, Nguyen et al. presented non-binary Varshamov-Tenengolts~(VT) codes that are capable of correcting a single deletion or a single insertion with linear time encoding-decoding algorithms. For more error-correction coding schemes with efficient decoding algorithms for DNA data storage, see \cite{wu_1, wu_2, wu_3}.}
\par
Park et al. \cite{park_lee_no} proposed an iterative decoding algorithm based on a mapping table for a constrained code that addresses the $GC$-content and the maximum homopolymer length requirements. Their method achieves normalized code rate $0.9165$ and $45\%-55\%$ $GC$-content range, and their iterative encoding algorithm has a mapping table with $48$ $3$-tuple $4$-ary entries as a building block. Liu et al. \cite{liu_he_tang} proposed a constrained coding scheme that achieves a $GC$-content in $[0.5 - \epsilon, 0.5+\epsilon]$ and eliminates homopolymers of length larger than $\ell$, and whose encoding-decoding algorithms have polynomial execution time and storage overhead. They also offered a coding scheme that satisfies a local $GC$-content constraint in order to further improve immunity against errors.

\subsection{Our Contribution and Organization of the Paper}

{ 
Our main result is designing the new D-LOCO codes for DNA data storage with low-complexity encoding-decoding algorithms and with desirable properties such as
\begin{itemize} 
\item[--] capacity-achievability, 
\item[--] reconfigurability, 
\item[--] low error propagation,
\item[--] parallelism, 
\item[--] substitution error detection, and
\item[--] local $GC$-content balance.
\end{itemize}
}

In particular, we devise the general D-LOCO encoding-decoding rule, which is given in Theorem~\ref{thm_general} in Subsection~\ref{dec_ell5}, after we discuss an interesting special case in Theorem~\ref{thm_ell3} in Subsection~\ref{dec_ell3}. Theorem~\ref{thm_general} provides a one-to-one mapping from an index set to the D-LOCO code, which is the encoding, and a one-to-one demapping from the D-LOCO code to the index set, which is the decoding. The encoding-decoding rule we obtain provides us with simple, low-complexity encoding-decoding algorithms (see Section~\ref{sec_balance} for algorithms). In fact, the storage overhead of our encoding-decoding algorithms turns out to be drastically low compared with those given in the literature {(see Subsection~\ref{complexity}).} With this idea of encoding-decoding, one just needs a simple adder that converts an index (binary message) to a codeword and vice versa. This provides the pivotal advantage of reconfigurability of the code, which is as easy as reprogramming an adder via a set of multiplexers. This ease of reconfigurability allows one to adopt different coding schemes in accordance with the requirement of the DNA data storage system at different stages of its lifetime {(see Subsection~\ref{reconfigurability}).}
\par
We also offer systematic approaches for balancing and bridging. Our approach for balancing incurs a minimal rate penalty that vanishes as the code length increases, which means it is capacity-achievable. This balancing approach guarantees not only global $GC$-content balance, but also local $GC$-content balance since it picks each codeword from two possible options according to the running $AT-GC$ disparity. Furthermore, we suggest various bridging schemes that enable error detection, which enhances system reliability.
\par
In Section~\ref{sec_card}, we introduce D-LOCO codes and study their cardinality. In Section~\ref{sec_dec}, we derive a simple encoding-decoding rule for D-LOCO codes forbidding runs of length higher than $3$ (Theorem~\ref{thm_ell3}) and then generalize this result to D-LOCO codes forbidding runs of length higher than any fixed value $\ell$ (Theorem~\ref{thm_general}). In Section~\ref{sec_bridge}, we propose four bridging schemes, three of which guarantee single substitution error detection. Assuming a $(1-p,p/3,p/3,p/3)$-symmetric DNA storage channel with substitution error rate $p$, a careful analysis of the probability of no-detection in the occurrence of multiple substitution errors is presented. In Section~\ref{sec_balance}, we provide the encoding-decoding algorithms and discuss how to balance the DNA sequence in order to achieve close percentage of $A$-$T$ and $G$-$C$ nucleotides overall. In Section~\ref{sec_rate}, we compare code rates at finite lengths when different bridging schemes are applied, and we show that D-LOCO codes are capacity-achieving. We conclude the paper with the complexity of encoding-decoding algorithms and other desirable properties of our proposed coding scheme.

%%%%%%%%%%%%%%%%%%%%%%%%%%%%%%%%%%%%%
\section{Definition and Cardinality}\label{sec_card}

In this section, we introduce D-LOCO codes and study their cardinality. 

\begin{definition} \label{D-LOCO_dna} (\hspace{-0.01em}\cite[Definition~1]{ahh_general}) For integers $\ell \geq 1$, the D-LOCO code $\mathcal{D}_{m,\ell}$ is defined as the set of all codewords of length $m$ defined over the alphabet $\{A, T, G, C\}$ that do not contain any pattern in $\mathcal{F}=\{\mathbf{\Lambda}^{\ell+1} \,|\, \Lambda \in \{A,T,C,G\}\}$. Here, $\mathbf{\Lambda}^{\ell+1}$ is the sequence of length $\ell+1$ all of whose symbols are $\Lambda$, and such a sequence $\mathbf{\Lambda}^{\ell+1}$ is simply called a run of length $\ell+1$. 

Elements of $\mathcal{D}_{m,\ell}$ are also called $\mathcal{F}$-constrained sequences of length $m$, and they are ordered lexicographically.
\end{definition}

Codewords in $\mathcal{D}_{m,\ell}$ are ordered in an ascending manner by following the rule $A < T < G < C$ for any symbol, and the symbol significance reduces from left to right. For more illustration, consider the two arbitrary codewords $\bold{c}$ and $\bold{c}'$ in $\mathcal{D}_{m,\ell}$. We say $\bold{c} < \bold{c}'$ if and only if for the first symbol position the two codewords differ at, $\bold{c}$ has a ``less'' symbol than that of $\bold{c}'$. This is how we define lexicographic ordering.

\begin{notation} The cardinality, i.e., codebook size, of the D-LOCO code $\mathcal{D}_{m,\ell}$ is denoted by $N_{\mathrm{D}}(m,\ell)$. However and for the ease of notation, we denote this cardinality by $N(m)$ whenever the context clarifies $\ell$.
\end{notation}

\begin{example} $\mathcal{D}_{m,m-1}$ is the set of all non-constant sequences of length $m$ over the alphabet $\{A,T,G,C\}$, and thus $N_{\mathrm{D}}(m,m-1)=4^m-4$; whereas $\mathcal{D}_{m,1}$ consists of all sequences of length $m$ whose consecutive terms are distinct, which means $N_{\mathrm{D}}(m,1)=4 \cdot 3^{m-1}$. Throughout the paper, $\mathcal{D}_{m,3}$ will be of special interest to us based on the literature on the topic (see Subsection~\ref{dec_ell3}) \cite{song_cai_etal, park_lee_no}.
\end{example} 

The D-LOCO code $\mathcal{D}_{m,\ell}$ is partitioned into 4 groups, consisting of total $4\ell$ subgroups, based on the set $\mathcal{F}$ of forbidden patterns. In particular, for $1 \leq k \leq \ell$, we have the following subgroups: \\
\indent \textbf{Subgroup A(k):} Codewords starting with $\mathbf{A}^{k}\Lambda$, where $\Lambda \in \{T,G,C\}$ from the left, \\
\indent \textbf{Subgroup T(k):} Codewords starting with $\mathbf{T}^{k}\Lambda$, where $\Lambda \in \{A,G,C\}$ from the left, \\
\indent \textbf{Subgroup G(k):} Codewords starting with $\mathbf{G}^{k}\Lambda$, where $\Lambda \in \{A,T,C\}$ from the left, and \\
\indent \textbf{Subgroup C(k):} Codewords starting with $\mathbf{C}^{k}\Lambda$, where $\Lambda \in \{A,T,G\}$ from the left. 

This partition of the D-LOCO code $\mathcal{D}_{m,\ell}$ is essential in both deriving its cardinality and the encoding-decoding rule. Let us first derive the cardinality $N(m)$ of $\mathcal{D}_{m,\ell}$ as a linear combination of cardinalities of D-LOCO codes with lengths smaller than $m$ using the group structure above. 

{We now introduce three notations and illustrate their relations to each other below.

\begin{notation} For $1 \leq k \leq \ell$ and $\ell < m$, let 
\begin{itemize} 

\item[1.] $N_{\Lambda_1,\Lambda_2, k}(m)$ denote the cardinality of the set of codewords in $\mathcal{D}_{m,\ell}$ starting with $\mathbf{\Lambda_1}^{k}\Lambda_2$, for $\Lambda_2 \in \{A,T,G,C\} \setminus \{\Lambda_1\}$, and let

\item[2.] $N_{\Lambda_1,k}(m)$ denote the number of codewords $\mathbf{c}=c_{m-1}c_{m-2}\dots c_1 c_0$ in $\mathcal{D}_{m,\ell}$ whose first $k$ symbols from the left are all $\Lambda_1$ and $c_{m-k-1} \neq \Lambda_1$. 

Note that we have
\begin{align} \label{not1} 
N_{\Lambda_1,k}(m) &= \sum_{\Lambda_2 \in \{A,T,G,C\} \setminus \{\Lambda_1\}} N_{\Lambda_1,\Lambda_2,k}(m),
\end{align}
for $1 \leq k \leq \ell$. Moreover, let 

\item[3.] $N_{\Lambda_1}(m)$ denote the number of codewords in $\mathcal{D}_{m,\ell}$ with $c_{m-1}=\Lambda_1$. Then, we have
\begin{equation} \label{not2} N_{\Lambda_1}(m) = \sum^{\ell}_{k=1} N_{\Lambda_1,k}(m). 
\end{equation}
\end{itemize}
\end{notation}
}
\begin{proposition} \label{prop_card} (\hspace{-0.01em}\cite[Equation~1]{immink_cai}) The cardinality $N(m)$ of the D-LOCO code $\mathcal{D}_{m,\ell}$, where $\ell \geq 1$, satisfies the following recursive relation for $m \geq \ell$: 
\begin{equation} \label{card1} N(m) = 3N(m-1)+3N(m-2)+\dots+3N(m-\ell).
\end{equation}
For $0 \leq m \leq \ell$, $$N(0) \triangleq \frac{4}{3}, \textrm{ and } N(m)=4^m \textrm{ for } 1 \leq m \leq \ell.$$
\end{proposition}

\begin{IEEEproof} We first consider the case of $m > \ell$. Combining (\ref{not1}) and (\ref{not2}) with the observation that 
\begin{align} N_{\Lambda_1,\Lambda_2,k}(m)=	N_{\Lambda_2}(m-k),
\end{align} 
we obtain  
\begin{align}\label{eq_symm} N_A(m) &\overset{(\ref{not2})}{=}\sum^{\ell}_{k=1} N_{A,k}(m)  \nonumber \\
& \overset{(\ref{not1})}{=} \sum^{\ell}_{k=1} [N_{A,T,k}(m)+N_{A,G,k}(m)+N_{A,C,k}(m)] \nonumber \\
& = \sum^{\ell}_{k=1} [N_T(m-k) +N_G(m-k) \nonumber  \\
& \hspace{1,5cm} +N_C(m-k)]. 
\end{align}
Thus, for all $\Lambda_1 \in \{A,T,G,C\}$, we have 
\begin{align} & N_{\Lambda_1}(m) = \sum^{\ell}_{k=1} \,\, \sum_{\Lambda_2 \in \{A,T,G,C\} \setminus \{\Lambda_1\}} N_{\Lambda_2}(m-k).
\end{align}
Adding these up over the $4$-ary alphabet, we obtain
\begin{align} N(m) &=N_A(m)+N_T(m)+N_G(m)+N_C(m) \nonumber \\
&=  3 \sum^{\ell}_{k=1} [N_A(m-k)+N_T(m-k) \nonumber \\
& \hspace{+1,3cm} +N_G(m,k)+N_C(m-k)] \nonumber \\
&= 3 \sum^{\ell}_{k=1} N(m-k),
\end{align}
proving the relation (\ref{card1}) for $m > \ell$. For $1 \leq m \leq \ell$, $\mathcal{D}_{m,\ell}$ is the set of all codewords in $\{A,T,G,C\}$ of length $m$, and thus $N(m)=4^m$. Moreover, the relation (\ref{card1}) holds in case $m = \ell$ by setting $N(0)=4/3$ to complete the summation $N(\ell) = 3 \sum_{i=1}^{\ell-1} 4^{\ell-i}+4 = 4^{\ell}$.
\end{IEEEproof}

%%%%%%%%%%%%%%%%%%%%%%%%%%%%%%%%%%%%%
\section{D-LOCO Encoding-Decoding Rule}\label{sec_dec}

\subsection{Encoding-Decoding Rule for $\ell=3$} \label{dec_ell3}

Now, we derive a formula that relates the lexicographic index of a D-LOCO codeword in the codebook to the codeword symbols. We call this formula the {\textit{encoding-decoding rule}} of D-LOCO codes since it is the foundation of the D-LOCO encoding and decoding algorithms illustrated by Examples~\ref{examp_ell3}, \ref{examp_ell}, and \ref{illustrate} below.

{Below, we study the encoding-decoding rule: $$g: \mathcal{D}_{m,\ell} \rightarrow \{0,1,\dots,N_{\mathrm{D}}(m,\ell)-1\}$$ that gives the index of any codeword in the codebook of $\mathcal{D}_{m,\ell}$ that is ordered lexicographically. 
\par 
In order to find the {\textit{index}} $g(\mathbf{c})$ of a codeword $\mathbf{c}=c_{m-1}c_{m-2}\dots c_1c_0$, we first study the contribution of each symbol $c_i$ (for $0 \leq i \leq m-1$) to $g(\mathbf{c})$. This contribution is denoted by $g_i(c_i)$, and we have 
\be g(\mathbf{c})=\sum^{m-1}_{i=0} g_i(c_i). \ee 
In \cite{ahh_general}, $g_i(c_i)$ is formulated as follows:
\be g_i(c_i)=\sum_{c_i' < c_i} N_{\mathrm{symb}}(m,c_{m-1}c_{m-2}\dots c_{i+1} c_i'), \ee
where $N_{\mathrm{symb}}(m,c_{m-1}c_{m-2}\dots c_{i+1} c_i')$ is the number of all codewords in $\mathcal{D}_{m,\ell}$ that start with the sequence $\mathbf{s}=c_{m-1}c_{m-2}\dots c_{i+1} c_i'$ from the left. We will compute $N_{\mathrm{symb}}(m,c_{m-1}c_{m-2}\dots c_{i+1} c_i')$ by counting the codewords $\mathbf{d}=c_i'd_{i-1}\dots d_1d_0$ in $\mathcal{D}_{i+1,\ell}$ such that $c_{m-1}c_m \dots c_{i+1}c_i'd_{i-1}\dots d_1d_0$ is a codeword in $\mathcal{D}_{m,\ell}$. For ease of expression, $c_{m-1}c_m \dots c_{i+1}c_i'd_{i-1}\dots d_1d_0$ will be called the \textit{wedge} of $\mathbf{s}$ and $\mathbf{d}$, and the operation itself will be called \textit{wedging}. 
}

Throughout the remaining part of this subsection, we set $\ell=3$. Here, $N_{\Lambda}(m)$ denotes the number of all codewords in $\mathcal{D}_{m,3}$ whose first letter is $\Lambda$, and $N_{\Lambda,k}(m)$ (for $1 \leq k \leq 3$) denotes the number of all codewords $\mathbf{c}=c_{m-1}c_{m-2}\dots c_1c_0$ in $\mathcal{D}_{m,3}$ whose first $k$ letters are all $\Lambda$ and $c_{m-k-1} \neq \Lambda$.~Note that using (\ref{eq_symm}) and the intrinsic symmetry of the code ($N_{A,k}(m)=N_{T,k}(m)=N_{G,k}(m)=N_{C,k}(m)$), we have the relation 
\begin{align}\label{eqn_3to4} 3N(m-k)=4N_{\Lambda,k}(m), \hspace{0.5cm} \textrm{ for } 1 \leq k \leq 3.
\end{align} 

{
\begin{remark} \label{out_of_codeword}
For Theorems \ref{thm_ell3}--\ref{thm_mostgeneral}, we define $c_i \triangleq \zeta$, for all $i > m-1$, to represent ``out of codeword bounds'' (see \cite{ahh_general}).
\end{remark}
}

\begin{theorem} \label{thm_ell3} The encoding-decoding rule $g: \mathcal{D}_{m,3} \rightarrow \{0,1,\dots,N(m)-1\}$ is as follows:
\be \label{eqn_rule3} g(\mathbf{c})=  \frac{3}{4} \sum^{m-1}_{i=0} \sum^{3}_{j=1} \sum^{j}_{k=1} (\mathsf{a}_{i,k}+\mathsf{t}_{i,k}+\mathsf{g}_{i,k}) N(i+j-3), \ee
where $N(0) \triangleq 4/3$, $N(-1) \triangleq 0$, and $N(-2) \triangleq 0$. Moreover, for all $\Pi > \Delta$ and for each $\Delta \in \{A, T, G\}$,
\begin{align} \delta_{i,1} &=1 \textrm{ only if } c_i=\Pi \textrm{ and } c_{i+1} \neq \Delta,\nonumber \\
\delta_{i,2} &=1 \textrm{ only if } c_{i+1}c_i=\Delta \Pi \textrm{ and } c_{i+2} \neq \Delta, \nonumber \\
\delta_{i,3} &=1 \textrm{ only if } c_{i+2}c_{i+1}c_i=\Delta\Delta \Pi \textrm{ and } c_{i+3} \neq \Delta,
\end{align} 
and in all other cases, $\delta_{i,k}=0$ for $i \in \{0,1,\dots,m-1\}$ and $k \in \{1,2,3\}$. \\
Here, $\Pi > \Delta$ is according to the lexicographic ordering rule, and $\delta$ stands for the small letter in $\{\mathsf{a},\mathsf{t},\mathsf{g}\}$ corresponding to $\Delta$ in $\{A,T,G\}$.
\end{theorem}
\begin{IEEEproof}
We study the symbol contributions $g_i(c_i)$ for each symbol $c_i$ in the alphabet $\{A,T,G,C\}$: \\
\textbf{Case 1:} If $c_i=A$, it is clear that $g_i(A) =0$.  \\
\textbf{Case 2:} If $c_i=T$, then 
\begin{align} g_i(T)= N_{\mathrm{symb}}(m,c_{m-1}c_{m-2}\dots c_{i+1} A).
\end{align}
\textbf{Case 2(a):} If $c_{i+1} \neq A$, then $\mathbf{s}=c_{m-1}c_{m-2}\dots c_{i+1} A$ can be wedged with any codeword in $\mathcal{D}_{i+1,3}$ which starts with $A$ from the left. In other words and using (\ref{eqn_3to4}),
\begin{align} & {N_{\mathrm{symb}}(m,c_{m-1}c_{m-2}\dots c_{i+1} A)} \nonumber \\ 
& \hspace{1,5em} = N_A(i+1) \nonumber \\
& \hspace{1,5em} = N_{A,1}(i+1)+N_{A,2}(i+1)+N_{A,3}(i+1) \nonumber \\
& \hspace{1,5em} =  \frac{3}{4}[N(i)+N(i-1)+N(i-2)].
\end{align}
\textbf{Case 2(b):} If $c_{i+1} = A$ but $c_{i+2} \neq A$, then $\mathbf{s}=c_{m-1}c_{m-2}\dots c_{i+2} AA$ can be wedged with any codeword  in $\mathcal{D}_{i+1,3}$ belonging to Subgroup A(1) or Subgroup A(2). Thus,
 \begin{align}
& {N_{\mathrm{symb}}(m,c_{m-1}c_{m-2}\dots c_{i+2} AA)}   \nonumber \\
& \hspace{2,5cm} = N_{A,1}(i+1)+N_{A,2}(i+1) \nonumber \\
& \hspace{2,5cm} = \frac{3}{4}[N(i)+N(i-1)].
\end{align} 
\textbf{Case 2(c):} If $c_{i+1} = c_{i+2} = A$ but $c_{i+3} \neq A$, then $\mathbf{s}=c_{m-1}c_{m-2}\dots c_{i+3} AAA$ can be wedged with any codeword in $\mathcal{D}_{i+1,3}$ belonging to Subgroup A(1), i.e.,
\begin{align} & N_{\mathrm{symb}}(m,c_{m-1}c_{m-2}\dots c_{i+3} AAA) \nonumber \\
& \hspace{1cm} = N_{A,1}(i+1) = \frac{3}{4}N(i). 
\end{align}
Since these subcases are all disjoint, we simply have
\begin{align}
g_i(T) &=N_{\mathrm{symb}}(m,c_{m-1}c_{m-2}\dots c_{i+1}A) \nonumber \\
&=\frac{3}{4}\bigg[ (\mathsf{a}_{i,1}+\mathsf{a}_{i,2}+\mathsf{a}_{i,3})N(i) \nonumber \\
& \hspace{1cm} +(\mathsf{a}_{i,1}+\mathsf{a}_{i,2})N(i-1)+\mathsf{a}_{i,1}N(i-2) \bigg] \nonumber \\
&= \frac{3}{4}\sum^{3}_{j=1} \sum^{j}_{k=1} \mathsf{a}_{i,k} N(i+j-3). 
\end{align} 
where $N(0) \triangleq 4/3$, $N(-1) \triangleq 0$, and $N(-2) \triangleq 0$. Moreover,
$\mathsf{a}_{i,1} =1 $ only if $c_i=T$  and $c_{i+1} \neq  A$, \\
$\mathsf{a}_{i,2} =1 $ only if $c_{i+1}c_i=AT$ and $c_{i+2} \neq A$, \\ 
$\mathsf{a}_{i,3} =1 $ only if $c_{i+2}c_{i+1}c_i=AAT$ and $c_{i+3} \neq A$, and \\
$\mathsf{a}_{i,k}=0$ otherwise.  \\
\textbf{Case 3:} If $c_i=G$, then 
\begin{align} g_i(G) &= N_{\mathrm{symb}}(m,c_{m-1}c_{m-2}\dots c_{i+1} A) \nonumber \\
&+ N_{\mathrm{symb}}(m,c_{m-1}c_{m-2}\dots c_{i+1} T).
\end{align} We study the term $N_{\mathrm{symb}}(m,c_{m-1}c_{m-2}\dots c_{i+1} T)$ in various cases below. \\
\textbf{Case 3(a):} If $c_{i+1} \neq T$, then $\mathbf{s}=c_{m-1}c_{m-2}\dots c_{i+1} T$ can be wedged with any codeword in $\mathcal{D}_{i+1,3}$ which starts with $T$. In other words,
\begin{align} & {N_{\mathrm{symb}}(m,c_{m-1}c_{m-2}\dots c_{i+1} T)} \nonumber \\
& \hspace{1cm} = N_T(i+1) \nonumber \\
& \hspace{1cm} = N_{T,1}(i+1)+N_{T,2}(i+1)+N_{T,3}(i+1) \nonumber \\
& \hspace{1cm} =  \frac{3}{4}[N(i)+N(i-1)+N(i-2)].
\end{align}
\textbf{Case 3(b):} If $c_{i+1} = T$ but $c_{i+2} \neq T$, then $\mathbf{s}=c_{m-1}c_{m-2}\dots c_{i+2} TT$ can be wedged with any codeword  in $\mathcal{D}_{i+1,3}$ belonging to Subgroup T(1) or Subgroup T(2). Thus,
 \begin{align}
& {N_{\mathrm{symb}}(m,c_{m-1}c_{m-2}\dots c_{i+2} TT)} \nonumber \\
& \hspace{1cm} = N_{T,1}(i+1)+N_{T,2}(i+1) \nonumber \\
& \hspace{1cm} = \frac{3}{4}[N(i)+N(i-1)].
\end{align} 
\textbf{Case 3(c):} If $c_{i+1} = c_{i+2} = T$ but $c_{i+3} \neq T$, then $\mathbf{s}=c_{m-1}c_{m-2}\dots c_{i+3} TTT$ can be wedged with any codeword in $\mathcal{D}_{i+1,3}$ belonging to Subgroup T(1), i.e.,
\begin{align} & {N_{\mathrm{symb}}(m,c_{m-1}c_{m-2}\dots c_{i+3} TTT)} \nonumber \\
& \hspace{1cm} = N_{T,1}(i+1) = \frac{3}{4}N(i). 
\end{align}
Since these subcases are all disjoint, we simply have the expression in (\ref{g_i(G)}), where $N(0) \triangleq 4/3$. Moreover, \\
$\mathsf{a}_{i,1} =1 $ only if $c_i = G$ and $c_{i+1} \neq  A$,\\
$\mathsf{a}_{i,2} =1 $ only if $c_{i+1}c_i=AG$ and $c_{i+2} \neq A$, \\ 
$\mathsf{a}_{i,3} =1 $ only if $c_{i+2}c_{i+1}c_i=AAG$  and $c_{i+3} \neq A$, and \\
$\mathsf{a}_{i,k}=0$ otherwise; in addition, \\
$\mathsf{t}_{i,1} =1 $ only if $c_i = G$ and $c_{i+1} \neq  T$,\\
$\mathsf{t}_{i,2} =1 $ only if $c_{i+1}c_i=TG$ and $c_{i+2} \neq T$, \\ 
$\mathsf{t}_{i,3} =1 $ only if $c_{i+2}c_{i+1}c_i=TTG$ and $c_{i+3} \neq T$, and \\
$\mathsf{t}_{i,k}=0$ otherwise. \\
\textbf{Case 4:} One can similarly study {the symbol contribution $g_i(c_i)$ for $c_i=C$} and find the number $N_{\mathrm{symb}}(m,c_{m-1}c_{m-2}\dots c_{i+1} G)$ of codewords in $\mathcal{D}_{m,3}$ that start with $\mathbf{s}=c_{m-1}c_{m-2}\dots c_{i+1} G$ {as well} to obtain $g_i(C)$.

Adding all $g_i(c_i)$'s up, we obtain the encoding-decoding rule as given in the theorem statement. 
\end{IEEEproof}

\begin{remark} \label{rem_alterexp} Note that we have stated the encoding-decoding rule in Theorem~\ref{thm_ell3} such that it aligns with its generalizations in the next section and in the Appendix. However, in our examples, we will use it in the form (\ref{alter_exp}) where $\delta_{i,k}$'s are as in the statement of Theorem~\ref{thm_ell3}. Observe that (\ref{eqn_rule3}) emerges directly from substituting (\ref{card1}) for $N(i+1)$ in (\ref{alter_exp}).
\end{remark}

\begin{figure*}[ht!]
\noindent\makebox[\linewidth]{\rule{\textwidth}{1,5pt}} \\
\begin{align} \label{g_i(G)}
g_i(G) &=N_{\mathrm{symb}}(m,c_{m-1}c_{m-2}\dots c_{i+1}A)+ N_{\mathrm{symb}}(m,c_{m-1}c_{m-2}\dots c_{i+1} T)  \nonumber \\
&=\frac{3}{4}\bigg[ (\mathsf{a}_{i,1}+\mathsf{t}_{i,1}+\mathsf{a}_{i,2}+t_{i,2}+\mathsf{a}_{i,3}+\mathsf{t}_{i,3})N(i) \nonumber \\
& \hspace{4cm} +(\mathsf{a}_{i,1}+\mathsf{t}_{i,1}+\mathsf{a}_{i,2}+t_{i,2})N(i-1)+(\mathsf{a}_{i,1}+\mathsf{t}_{i,1})N(i-2) \bigg] \nonumber \\
&= \frac{3}{4}\sum^{3}_{j=1} \sum^{j}_{k=1} (\mathsf{a}_{i,k}+\mathsf{t}_{i,k}) N(i+j-3). 
\end{align}
\noindent\makebox[\linewidth]{\rule{\textwidth}{1,5pt}}
\end{figure*}
\begin{figure*}[ht!]
\begin{align} \label{alter_exp} g(\mathbf{c}) &=\sum^{m-1}_{i=0} \bigg[ \frac{\mathsf{a}_{i,1}+\mathsf{t}_{i,1}+\mathsf{g}_{i,1}}{4}N(i+1)+\frac{\mathsf{a}_{i,2}+t_{i,2}+\mathsf{g}_{i,2}}{4}(N(i+1)-3N(i-2)) \nonumber \\
& \hspace{5cm}+\frac{3(\mathsf{a}_{i,3}+\mathsf{t}_{i,3}+\mathsf{g}_{i,3})}{4}N(i) \bigg]. 
\end{align}
\noindent\makebox[\linewidth]{\rule{\textwidth}{1,5pt}}
\end{figure*}

\begin{example} \label{examp_ell3} Consider the encoding-decoding rule $$g: \mathcal{D}_{4,3} \rightarrow  \{0,1,\dots,251\},$$ where $N(-2) \triangleq 0$, $N(-1) \triangleq 0$, $N(0) \triangleq 4/3$, $N(1) = 4$, $N(2) = 16$, $N(3) = 64$, and $N(4)=4^4-4=252$. We discuss some instances and provide a detailed explanation for the codeword $ATGC$:
\begin{itemize} 
\item For the codeword $\mathbf{c}=AAAT \in \mathcal{D}_{4,3}$, we have $\mathsf{a}_{i,k}=\mathsf{t}_{i,k}=\mathsf{g}_{i,k}=0$ for all $i \in \{0,1,2,3\}$ and $k \in \{1,2,3\}$. Therefore, $$g(AAAT)=g_0(T)=0.$$
\item For the codeword $\mathbf{c}=ATAT \in \mathcal{D}_{4,3}$, we have $\mathsf{a}_{2,2}=\mathsf{a}_{0,2}=1$ and all other variables are zero. Therefore, 
\begin{align*} g(ATAT) &= g_2(T)+g_0(T) \\
&= \frac{1}{4}(N(3)-3N(0)) \\
&+\frac{1}{4}(N(1)-3N(-2))= 15+1=16.
%&g(ATAT)=g_0(T)+g_2(T) \\
%& \hspace{0.5em} =\frac{3}{4}(N(0)+N(-1))+\frac{3}{4}(N(2)+N(1)) \\
%& \hspace{0.5em} = 1+15=16,
\end{align*}
\item For the codeword $\mathbf{c}=ATGC \in \mathcal{D}_{4,3}$, we have $\mathsf{a}_{2,2}=\mathsf{a}_{1,1}=\mathsf{t}_{1,2}=\mathsf{a}_{0,1}=\mathsf{t}_{0,1}=\mathsf{g}_{0,2}=1$ and all other variables are zero. {Note that by (\ref{alter_exp}),
\begin{itemize} 
\item[1.] $g_3(A) = 0$,
\item[2.] As $c_3c_2= AT$ and $c_4 = \zeta$, $\mathsf{a}_{2,2}=1$ so that
$$\displaystyle g_2(T) = \frac{\mathsf{a}_{2,2}}{4}(N(3)-N(0)),$$
\item[3.] As $c_2c_1=TG$ and $c_3 \neq T$, $\mathsf{a}_{1,1}=\mathsf{t}_{1,2}=1$ so that
$$\displaystyle g_1(G)= \frac{\mathsf{a}_{1,1}}{4}N(2)+\frac{\mathsf{t}_{1,2}}{4}(N(2)-N(-1)), \textrm{ and}$$
\item[4.] As $c_1c_0=GC$ and $c_2 \neq G$, $\mathsf{a}_{0,1}=\mathsf{t}_{0,1}=\mathsf{g}_{0,2}=1$ so that
$$\displaystyle g_0(C)= \frac{\mathsf{a}_{0,1}+\mathsf{t}_{0,1}}{4}N(1)+\frac{\mathsf{g}_{0,2}}{4}(N(1)-N(-2)).$$
\end{itemize}
}
Therefore,
\begin{align*} g(ATGC) &= \frac{1}{4}(N(3)-3N(0)) \\
&+\frac{1}{4}N(2)+\frac{1}{4}(N(2)-3N(-1)) \\
&+\frac{2}{4}N(1)+\frac{1}{4}(N(1)-3N(-2)) \\
&= 15+4+4+2+1 = 26.
%&g(ATGC)  \\
%& \hspace{0.5em} =g_0(C)+g_1(G)+g_2(T)  \\
%& \hspace{0.5em} = \frac{3}{4}(3N(0)+N(-1))+\frac{3}{4}(2N(1)+N(0)) \\
%& \hspace{0.5em} +\frac{3}{4}(N(2)+N(1)) \\
%& \hspace{0.5em} =  3+7+15 = 26.
\end{align*}
\item For the codeword $\mathbf{c}=GGGC \in \mathcal{D}_{4,3}$, we have $\mathsf{a}_{3,1}=\mathsf{t}_{3,1}=\mathsf{a}_{2,1}=\mathsf{t}_{2,1}=\mathsf{a}_{1,1}=\mathsf{t}_{1,1}=\mathsf{a}_{0,1}=\mathsf{t}_{0,1}=1$ and all other variables are zero. Therefore, 
\begin{align*} g(GGGC) &= \frac{2}{4}N(4)+\frac{2}{4}N(3) \\
&+\frac{2}{4}N(2) + \frac{2}{4}N(1)= 168.
%& g(GGGC)  \\
%& \hspace{0.5em} =g_0(C)+g_1(G)+g_2(G)+g_3(G)  \\
%& \hspace{0.5em} =\frac{3}{4}(2N(0)+2N(-1)+2N(-2)) \\
%& \hspace{0.5em} +\frac{3}{4}(2N(1)+2N(0)+2N(-1)) \\
%& \hspace{0.5em} +\frac{3}{4}(2N(2)+2N(1)+2N(0)) \\
%& \hspace{0.5em} +\frac{3}{4}(2N(3)+2N(2)+2N(1)) \\
%& \hspace{0.5em} = 2+8+32+126=168,
\end{align*}
\item For the codeword $\mathbf{c}=CCCG \in \mathcal{D}_{4,3}$, we have $\mathsf{a}_{3,1}=\mathsf{t}_{3,1}=\mathsf{g}_{3,1}=\mathsf{a}_{2,1}=\mathsf{t}_{2,1}=\mathsf{g}_{2,1}=\mathsf{a}_{1,1}=\mathsf{t}_{1,1}=\mathsf{g}_{1,1}=\mathsf{a}_{0,1}=\mathsf{t}_{0,1}=1$ and all other variables are zero. Therefore, 
\begin{align*} g(CCCG) &= \frac{3}{4}N(4)+\frac{3}{4}N(3)+\frac{3}{4}N(2)+\frac{2}{4}N(1) \\
&= 189+48+12+2 =251.
%& g(CCCG) \\
%& \hspace{0.5em} =g_0(G)+g_1(C)+g_2(C)+g_3(C) \\
%& \hspace{0.5em} =\frac{3}{4}(2N(0)+2N(-1)+2N(-2)) \\
%& \hspace{0.5em} +\frac{3}{4}(3N(1)+3N(0)+3N(-1)) \\
%& \hspace{0.5em} +\frac{3}{4}(3N(2)+3N(1)+3N(0)) \\
%& \hspace{0.5em} +\frac{3}{4}(3N(3)+3N(2)+3N(1)) \\
%& \hspace{0.5em} =2+12+48+189 =251.
\end{align*}
\end{itemize}
\end{example}

\subsection{Encoding-Decoding Rule for General $\ell$} \label{dec_ell5}

Below, we study the symbol contributions of each symbol for general $\ell \geq 1$, and we obtain a formula (in terms of cardinalities $N(i)$) for the encoding-decoding rule $g: \mathcal{D}_{m,\ell} \rightarrow \{0,1,\dots,N(m)-1\}$ that gives the index of codewords in $\mathcal{D}_{m,\ell}$. Recall that $N_{\Lambda}(m)$ denotes the number of all codewords in $\mathcal{D}_{m,\ell}$ whose first letter from the left is $\Lambda$, and $N_{\Lambda,k}(m)$ (for $1 \leq k \leq l$) denotes the number of all codewords in $\mathcal{D}_{m,\ell}$ whose first $k$ letters are $\Lambda$ and $c_{m-k-1} \neq \Lambda$. Moreover, we have the relation $3N_{\Lambda}(m-k)=4N_{\Lambda,k}(m)$ for $k \geq 1$.

\begin{theorem} \label{thm_general} The encoding-decoding rule $g: \mathcal{D}_{m,\ell} \rightarrow \{0,1,\dots,N(m)-1\}$, for general $\ell \geq 1$, is as follows:
\be
g(\mathbf{c})= \frac{3}{4} \sum^{m-1}_{i=0} \sum^{\ell}_{j=1} \sum^{j}_{k=1} (\mathsf{a}_{i,k}+\mathsf{t}_{i,k}+\mathsf{g}_{i,k}) N(i+j-\ell), \label{eqn_rulegen}
\ee
\\
where $N(0) \triangleq 4/3$ and $N(-1) \triangleq \dots \triangleq N(2-\ell) \triangleq N(1-\ell) \triangleq 0$. Moreover, for all $\Pi > \Delta$ and for each $\Delta \in \{A, T, G\}$,
\begin{align} \label{coeffs_gen}
\delta_{i,1} &=1 \textrm{ only if } c_i=\Pi \textrm{ and } c_{i+1} \neq \Delta, \nonumber \\
\delta_{i,k} &=1 \textrm{ only if } c_{i+k-1}\dots c_{i+1}c_i=\mathbf{\Delta}^{k-1} \Pi \textrm{ and } c_{i+k} \neq \Delta,
\end{align}
where $2 \leq k \leq \ell$ (these are $\ell-1$ statements), and in all other cases, $\delta_{i,k}=0$ for $i \in \{1,2,\dots,m\}$ and $k \in \{1,2,\dots,\ell\}$. \\
Here, $\Pi > \Delta$ is according to the lexicographic ordering rule, and $\delta$ stands for the small letter in $\{\mathsf{a},\mathsf{t},\mathsf{g}\}$ corresponding to $\Delta$ in $\{A, T, G\}$.
\end{theorem}
\begin{IEEEproof}
{First, recall that $c_i \triangleq \zeta$ (a letter outside of the alphabet), for all $i > m-1$ (see Remark~\ref{out_of_codeword}).}

Symbol contributions are as follows: \\
\textbf{Case 1:} If $c_i=A$, it is clear that $g_i(A) =0$.  \\
\textbf{Case 2:} If $c_i=T$, then 
\begin{align} &g_i(T) = N_{\mathrm{symb}}(m,c_{m-1}c_{m-2}\dots c_{i+1}A) \nonumber \\
&=\mathsf{a}_{i,1}[N_{A,1}(i+1)+N_{A,2}(i+1) +\dots +N_{A,\ell}(i+1)] \nonumber \\
&+\mathsf{a}_{i,2}[N_{A,1}(i+1)+N_{A,2}(i+1)+\dots+N_{A,\ell-1}(i+1)] \nonumber \\
&+\dots \nonumber \\
&+\mathsf{a}_{i,\ell-1}[N_{A,1}(i+1)+N_{A,2}(i+1)]+\mathsf{a}_{i,\ell}N_{A,1}(i+1) \nonumber \\
&= (\mathsf{a}_{i,1}+\mathsf{a}_{i,2}+\dots+\mathsf{a}_{i,\ell})N_{A,1}(i+1) + \dots \nonumber \\
&+(\mathsf{a}_{i,1}+\mathsf{a}_{i,2})N_{A,\ell-1}(i+1)+\mathsf{a}_{i,1}N_{A,\ell}(i+1) \nonumber \\ 
&=\frac{3}{4} \sum^{\ell}_{j=1} \sum^{j}_{k=1} \mathsf{a}_{i,k} N(i+j-\ell),
\end{align}
where $N(0) \triangleq 4/3$ and $N(-1) \triangleq \dots \triangleq N(2-\ell) \triangleq N(1-\ell) \triangleq 0$. Moreover, \\
$\mathsf{a}_{i,1} =1 $ only if $c_i=A$ and $c_{i+1} \neq A$,\\
$\mathsf{a}_{i,k} =1 $ only if $c_{i+k-1}\dots c_{i+1}c_i=\mathbf{A}^{k-1} T$ and $c_{i+k} \neq A$ where $2 \leq k \leq \ell$, and \\ 
$\mathsf{a}_{i,k}=0$ otherwise. \\
\textbf{Case 3 and Case 4:} One can similarly study and find the numbers $N_{\mathrm{symb}}(m,c_{m-1}c_{m-2}\dots c_{i+1} T)$ and $N_{\mathrm{symb}}(m,c_{m-1}c_{m-2}\dots c_{i+1} G)$ as well to obtain $g_i(G)$ and $g_i(C)$.

Adding all $g_i(c_i)$'s up, we obtain the encoding-decoding rule as given in the theorem statement. 
\end{IEEEproof}

\begin{example} \label{examp_ell} Consider the encoding-decoding rule $$g: \mathcal{D}_{5,4} \rightarrow  \{0,1,\dots,1019\}$$ where $N(-3) \triangleq N(-2) \triangleq N(-1) \triangleq 0$, $N(0) \triangleq 4/3$, $N(1) = 4$, $N(2) = 16$, $N(3) = 64$, $N(4) = 256$, and $N(5)=4^5-4=1020$. We discuss some instances and provide a detailed explanation for the codeword $TAATT$:
\begin{itemize} 
\item For the codeword $\mathbf{c}=AAAAT \in \mathcal{D}_{5,4}$, we have $\mathsf{a}_{i,k}=\mathsf{t}_{i,k}=\mathsf{g}_{i,k}=0$ for all $i \in \{0,1,2,3,4\}$ and $k \in \{1,2,3,4\}$. Therefore, $$g(AAAAT)=g_0(T)=0.$$
\item For the codeword $\mathbf{c}=TAATT \in \mathcal{D}_{5,4}$, we have $\mathsf{a}_{4,1}=\mathsf{a}_{1,3}=\mathsf{a}_{0,1}=1$ and all other variables are zero. {Note that by (\ref{eqn_rulegen}), 
\begin{itemize} 
\item[1.] $g_3(A) = g_2(A) = 0$,
\item[2.] As $c_4= T$ and $c_5 = \zeta$, $\mathsf{a}_{4,1}=1$ so that
$$\displaystyle g_4(T) = \frac{3\mathsf{a}_{4,1}}{4}(N(4)+N(3)+N(2)+N(1)),$$
\item[3.] As $c_3c_2c_1 = AAT$ and $c_4 \neq A$, $\mathsf{a}_{1,3}=1$
$$\displaystyle g_1(T)= \frac{3\mathsf{a}_{1,3}}{4}(N(1)+N(0)), \textrm{ and}$$
\item[4.] As $c_1c_0=TT$ and $c_2 \neq T$, $\mathsf{a}_{0,1}=1$ so that
$$\displaystyle g_0(T)= \frac{3\mathsf{a}_{0,1}}{4}N(0).$$
\end{itemize}
}
Therefore, 
\begin{align*} &g(TAATT) \\
& \hspace{0.5em} =g_4(T)+g_1(T)+g_0(T) \\
& \hspace{0.5em} =\frac{3}{4}(N(4)+N(3)+N(2)+N(1)) \\
& \hspace{0.5em} +\frac{3}{4}(N(1)+N(0))+\frac{3}{4}N(0) \\
& \hspace{0.5em} = 255+4+1 = 260.
\end{align*}
\item For the codeword $\mathbf{c}=GGGGC \in \mathcal{D}_{5,4}$, we have $\mathsf{a}_{4,1}=\mathsf{t}_{4,1}=\mathsf{a}_{3,1}=\mathsf{t}_{3,1}=\mathsf{a}_{2,1}=\mathsf{t}_{2,1}=\mathsf{a}_{1,1}=\mathsf{t}_{1,1}=\mathsf{a}_{0,1}=\mathsf{t}_{0,1}=1$ and all other variables are zero. Therefore, 
\vspace{-0.1em}\begin{align*}  & g(GGGGC) \\
& \hspace{0.5em}=g_4(G)+g_3(G)+g_2(G)+g_1(G)+g_0(C) \\
& \hspace{0.5em}= \frac{3}{4}(2N(4)+2N(3)+2N(2)+2N(1)) \\
& \hspace{0.5em}+ \frac{3}{4}(2N(3)+2N(2)+2N(1)+2N(0)) \\
& \hspace{0.5em}+ \frac{3}{4}(2N(2)+2N(1)+2N(0)) \\
& \hspace{0.5em}+ \frac{3}{4}(2N(1)+2N(0)) + \frac{3}{4}(2N(0)) \\
& \hspace{0.5em}= 510+128+32+8+2=680.
\end{align*}
\item For the codeword $\mathbf{c}=CATGC \in \mathcal{D}_{5,4}$, we have $\mathsf{a}_{4,1}=\mathsf{t}_{4,1}=\mathsf{g}_{4,1}=\mathsf{a}_{2,2}=\mathsf{a}_{1,1}=\mathsf{t}_{1,2}=\mathsf{a}_{0,1}=\mathsf{t}_{0,1}=\mathsf{g}_{0,2}=1$ and all other variables are zero. Therefore, 
\begin{align*} & g(CATGC) \\
& \hspace{0.5em}= g_4(C)+g_2(T)+g_1(G)+g_0(C) \\
& \hspace{0.5em}= \frac{3}{4}(3N(4)+3N(3)+3N(2)+3N(1)) \\
& \hspace{0.5em}+ \frac{3}{4}(N(2)+N(1)+N(0)) \\
& \hspace{0.5em}+ \frac{3}{4}(2N(1)+2N(0))+\frac{3}{4}(3N(0)) \\
& \hspace{0.5em}= 765+16+8+3 = 792.
\end{align*}
\item For the codeword $\mathbf{c}=CCCCG \in \mathcal{D}_{5,4}$, we have $\mathsf{a}_{i,1}=\mathsf{t}_{i,1}=\mathsf{g}_{i,1}=1$, for all $i \in \{1,2,3,4\}$, $\mathsf{a}_{0,1}=\mathsf{t}_{0,1}=1$, and all other variables are zero. Therefore, 
\begin{align*} & g(CCCCG) \\
& \hspace{0.5em} = g_4(C)+g_3(C)+g_2(C)+g_1(C)+g_0(G) \\
& \hspace{0.5em} = \frac{3}{4}(3N(4)+3N(3)+3N(2)+3N(1)) \\
& \hspace{0.5em} + \frac{3}{4}(3N(3)+3N(2)+3N(1)+3N(0)) \\
& \hspace{0.5em} + \frac{3}{4}(3N(2)+3N(1)+3N(0)) \\
& \hspace{0.5em} + \frac{3}{4}(3N(1)+3N(0))+\frac{3}{4}(2N(0)) \\
& \hspace{0,5em} = 765+192+48+12+2 =1019.
\end{align*}
\end{itemize}
\end{example}

\begin{remark} We provide additional generalization of Theorem~\ref{thm_general} in the Appendix to cover $q$-ary constrained codes eliminating runs of length exceeding $\ell \geq 1$ for any $q \geq 2$ (see Theorem~\ref{thm_mostgeneral}).
\end{remark}

\begin{remark} The above examples illustrate the use of the encoding-decoding rule for the decoding procedure. In the next section, Example~\ref{illustrate} briefly illustrates the use of the encoding-decoding rule for the encoding procedure. For the encoding-decoding algorithms, see Section~\ref{sec_balance}.
\end{remark}

%\begin{example}
%For instance, we have (as verified)
%\begin{align*}
%g(AAAAT) &= 0 \\ 
%g(ACCCG) &= \frac{2}{4}N(1)+\frac{3}{4}N(2)+\frac{3}{4}N(3)+\frac{1}{4}(N(4)-3N(0))+\frac{2}{4}N(4) \\
%&= 2+12+48+63+128 =253 \\ 
%g(TAATT) &= \frac{1}{4}N(1)+\frac{3}{4}N(1)+\frac{3}{4}N(0)+\frac{1}{4}N(5) = 259 \\
%g(TATAT) &= \frac{1}{4}(N(1)-3N(-3))+\frac{1}{4}(N(3)-3N(-1))+\frac{1}{4}N(5)= 1+16+255=272 \\
%g(GGGGC) &= \frac{2}{4}N(1)+\frac{2}{4}N(2)+\frac{2}{4}N(3) + \frac{2}{4}N(4)+\frac{2}{4}N(5)= 680 \\
%g(CATGC) &= \frac{2}{4}N(1)+\frac{1}{4}(N(1)-3N(-3))+\frac{1}{4}N(2)+\frac{1}{4}(N(2)-3N(-2))\\
%&+\frac{1}{4}(N(3)-3N(-1))+\frac{3}{4}N(5) \\
%&= 2+1+4+4+16+765 = 792
%\end{align*}
%\end{example}

%%%%%%%%%%%%%%%%%%%%%%%%%%%%%%%%%%%%%
\section{Bridging and Error Detection}\label{sec_bridge}

In this section, we present our bridging schemes. {Using a single codeword for the whole DNA strand results in higher encoding-decoding complexity and codeword-to-message error propagation \cite{ahh_loco, ahh_general}. Hence, it is essential while encoding data in a DNA strand to concatenate shorter codewords in order to prevent forbidden patterns at the transition between consecutive codewords. This process is called bridging.} Our bridging schemes do not allow same-symbol runs of length higher than $3$, i.e., they are designed for D-LOCO codes with $\ell \geq 3$. We note that these schemes are selected to also enable error detection (which will be discussed in this section) without increasing the $AT-GC$ disparity (which will be discussed in the next section).

\vspace{-0.3em}
\subsection{Bridging Scheme I} \label{bridging_scheme_1}
This is a one-symbol bridging scheme. Let $\mathbf{c}_1$ and $\mathbf{c}_2$ be two consecutive codewords in the DNA data stream, where $\mathbf{c}_1$ ends with the symbol $\Lambda_1$ and $\mathbf{c}_2$ starts with the symbol $\Lambda_2$. We can bridge the two codewords $\mathbf{c}_1$ and $\mathbf{c}_2$ by inserting a symbol from the set $\{A,T,G,C\} \setminus \{\Lambda_1,\Lambda_2\}$. Moreover, this single bridging symbol enables encoding $1$-bit of information as follows:
{\begin{itemize}
\item[(i)] To encode $0$ (from the input message stream), we pick the letter having the lowest lexicographic index in $\{A,T,G,C\} \setminus \{\Lambda_1,\Lambda_2\}$. 
\item[(ii)] To encode $1$, we pick the letter having the highest lexicographic index in $\{A,T,G,C\} \setminus \{\Lambda_1,\Lambda_2\}$.
\end{itemize}
}
\vspace{-0.2em}
\subsection{Bridging Scheme II}
We now discuss a simple three-symbol bridging scheme, which has two versions, to ensure single substitution error detection per codeword. Let $\mathbf{c}_1$ and $\mathbf{c}_2$ be two consecutive codewords, where $\mathbf{c}_1$ ends with the symbol $\Lambda_1$ and $\mathbf{c}_2$ starts with the symbol $\Lambda_2$. Let also $\Lambda_3$ denote the check-sum of $\mathbf{c}_1$, i.e., the check-sum of the symbols in $\mathbf{c}_1$, where this check-sum is computed according to the correspondence ($c_i \longleftrightarrow a_i$): 
\begin{align} \label{eqn_int} A &\longleftrightarrow 0\,(\mathrm{mod}\,4), \hspace{1cm} G \longleftrightarrow 2\,(\mathrm{mod}\,4), \nonumber \\
T &\longleftrightarrow 1\,(\mathrm{mod}\,4), \hspace{1cm} C \longleftrightarrow 3\,(\mathrm{mod}\,4).
\end{align}
We define the integer $(\mathrm{mod}\,4)$ equivalent of symbol $c_i$ according to (\ref{eqn_int}) as $a_i$. Note that $\Lambda_3$ will be updated shortly. Here, we bridge the two codewords $\mathbf{c}_1$ and $\mathbf{c}_2$ as follows:
{ 
\begin{itemize}
\item[(i)] Assign $\Lambda_3$ as the middle bridging symbol. 
\item[(ii)] Update the check-sum $\Lambda_3$ by adding $(\mathrm{mod}\,4)$ the integer representation of the extra two message stream bits to encode within bridging (equivalent to one extra letter in total). 
\item[(iii)] Set $\Lambda_4$ to the letter having the lowest lexicographic index in $\{A,T,G,C\} \setminus \{\Lambda_1,\Lambda_3\}$ if the first bit to encode is $0$. 
\item[(iv)] Set $\Lambda_4$ to the letter having the highest lexicographic index in $\{A,T,G,C\} \setminus \{\Lambda_1,\Lambda_3\}$ if the first bit to encode is $1$.
\item[(v)] Similarly, choose $\Lambda_5$ from the set $\{A,T,G,C\} \setminus \{\Lambda_2,\Lambda_3\}$ to encode the second message stream bit.
\end{itemize}
}
For clarity, the sequence order is $\Lambda_1 \ \Lambda_4 \Lambda_3 \Lambda_5 \ \Lambda_2$. This version is called Bridging Scheme~II-A. A different treatment of $\Lambda_5$ creates the second version of the scheme.
\par
If we instead choose $\Lambda_5$ as the letter having the highest lexicographic index in the set $\{A,T\} \setminus \{\Lambda_2\}$ in case $\Lambda_{3} \in \{G,C\}$ and in the set $\{G,C\} \setminus \{\Lambda_2\}$ in case $\Lambda_{3} \in \{A,T\}$, this version is called Bridging Scheme~II-B. This version has a balancing advantage (see Remark~\ref{rem_dispar} in Section~\ref{sec_balance} for more details). Observe that in this case, there is only $1$ extra bit (say $b$) encoded within bridging. Thus, $\Lambda_3$ is updated by adding $(\mathrm{mod}\,4)$ the integer representation of the binary $2$-tuple $b0$.
\par
We illustrate how to apply this bridging scheme in the following example: 
\begin{example} \label{illustrate}
We encode the $38$-bit binary message stream $$\mathbf{b}=1010.1000.1100.1111.1010.1010.1101.1010.0111.11$$ into codewords in $\mathcal{D}_{9,3}$ by applying Bridging Scheme~II-A ($b_i$ denotes the $i^{\textrm{th}}$ digit of $\mathbf{b}$, starting from $b_1$ at the left):
\begin{itemize}
\item From Proposition~\ref{prop_card}, $N(5) = 996$, $N(6) = 3936$, $N(7) = 15552$, $N(8) = 61452$, and $N(9) = 242820$.
\item Our rule is $g:\mathcal{D}_{9,3} \rightarrow \{0,1,\dots,242819\}$ in (\ref{alter_exp}), where the message length is $\lfloor \log_2(242819) \rfloor=17$. The first $17$ bits form a message that corresponds to $86431$ in decimal integer. We briefly illustrate how to obtain the codeword of index $86431$ via the D-LOCO encoding-decoding rule:
\begin{itemize} 
\item Initialize $\mathrm{residual}=86431$.  
\item At $i=8$, $\frac{1}{4}N(9) \leq \mathrm{residual} < \frac{2}{4}N(9)$. Consequently, $\mathsf{a}_8=1$. Update $\mathrm{residual} = 86431-\frac{1}{4}N(9)=25726$.
\item At $i=7$, $\frac{1}{4}(N(8)-N(5)) \leq \mathrm{residual} < \frac{2}{4}N(8)=30726$. Consequently, $\mathsf{a}_7=1$. Update $\mathrm{residual} = 25726-\frac{1}{4}(N(8)-N(5))=11110$.
\item At $i=6$, $\frac{2}{4}N(7) \leq \mathrm{residual} < \frac{3}{4}N(7)=11664$. Consequently, $\mathsf{a}_6=2$. Update $\mathrm{residual} = 11110-\frac{2}{4}N(7)=3334$.
\end{itemize}
After repeating this procedure until $i=0$, we finally obtain $${\mathbf{a}_1=112321323 \implies \mathbf{c}_1=TTGCGTCGC.}$$ The bits $b_{20}$--$b_{36}$ form a message that corresponds to $44455$ in decimal integer, and the codeword corresponding to index $44455$ is obtained similarly: 
$${\mathbf{a}_2=023300311 \implies \mathbf{c}_2=AGCCAACTT.}$$
\item Since the check-sum of the first codeword is $1+1+2+3+2+1+3+2+3 \text{ } \,(\mathrm{mod}\,4)=2$ and $$b_{18}b_{19}=01 \longleftrightarrow 1\,(\mathrm{mod}\,4),$$ the middle bridging symbol is $C$, corresponding to $3$.
\item Given that $b_{18}=0$ in the binary message stream, the first bridging symbol between $TTGCGTCG\mathbf{C}$ and the check-sum symbol $C$ is $A$ (which is the letter having the lowest lexicographic index in $\{A,T,G\}$).
\item Similarly, given that $b_{19}=1$ in the message stream, the third bridging symbol between $\mathbf{A}GCCAACTT$ and the check-sum symbol $C$ is $G$ (which is the letter having the highest lexicographic index in $\{T,G\}$).
\item By computing the check-sum of the second codeword and using $b_{37}b_{38}$, the last $3$ coded symbols become $CAC$.
\end{itemize}
Hence, the given binary message stream is encoded as $TTGCGTCGC\mathbf{ACG}AGCCAACTT\mathbf{CAC}$. Note that if we drop $b_{19}$ and $b_{38}$ from the message stream above and adopt Bridging Scheme~II-B, the resulting $36$-bit message stream is encoded as $TTGCGTCGC\mathbf{AGT}AGCCAACTT\mathbf{GCT}$.
\end{example}

{\begin{remark} Note that if the written $\mathbf{c}_1$ has a single substitution error (or if an error occurs on the check-sum symbol itself), then $\mathbf{c}_1$ and the check-sum $\Lambda_3$ will be inconsistent, and in this case, the error is detected. Similarly, in case there is an error at the first or the third bridging symbol, $\Lambda_4$ or $\Lambda_5$, it is again detected since the check-sum $\Lambda_3$ is computed by taking the extra information (two message stream bits) encoded in these bridging symbols into account. This verifies the error detection property for Bridging Scheme~II-A. 
\end{remark}
}

\begin{remark} Note that the bridging letters $\Lambda_4$ and $\Lambda_5$ are used to separate the check-sum symbol from the codewords in order for forbidden patterns not to show up at the transition. However, one symbol is the minimum we can allocate for the check-sum given the run-length constraint. In the next subsection, we show how using more symbols for the check-sum can reduce the probability of not detecting errors.
\end{remark}

\subsection{Bridging Scheme III} 
We now discuss a five-symbol bridging scheme that ensures single substitution error detection per codeword for a D-LOCO code forbidding runs of length at least $4$, i.e., $\ell=3$, and achieves lower probability of no-detection in case multiple errors occur. Let $\mathbf{c}$ and $\mathbf{d}$ in $\mathcal{D}_{m,3}$ be two consecutive codewords whose length is divisible by $3$, i.e., $m=3m'$ for some integer $m'$. Moreover, we have $\mathbf{c}=\mathbf{c}_{1}\mathbf{c}_{2}\mathbf{c}_{3}$. Suppose $\mathbf{c}$ ends with the symbol $\Lambda_1$ and $\mathbf{d}$ starts with the symbol $\Lambda_2$. Let also $\Lambda_{3,i}$ denote the check-sum of the codeword $\mathbf{c}_{i}$ for $i=1,2,3$. Here, we bridge the two codewords $\mathbf{c}$ and $\mathbf{d}$ as follows: 
{
\begin{itemize}
\item[(i)] Assign $\mathbf{\Lambda}_3 \triangleq \Lambda_{3,1}\Lambda_{3,2}\Lambda_{3,3}$ as the middle bridging pattern. 
\item[(ii)] Set $\Lambda_4$ to the letter having the highest lexicographic index in the set $\{A,T\} \setminus \{\Lambda_1\}$ in case $\Lambda_{3,1} \in \{G,C\}$ and in the set $\{G,C\} \setminus \{\Lambda_1\}$ in case $\Lambda_{3,1} \in \{A,T\}$. 
\item[(iii)] Set $\Lambda_5$ to the letter having the highest lexicographic index in the set $\{A,T\} \setminus \{\Lambda_2\}$ in case $\Lambda_{3,3} \in \{G,C\}$ and in the set $\{G,C\} \setminus \{\Lambda_2\}$ in case $\Lambda_{3,3} \in \{A,T\}$.
\end{itemize}
}
For clarity, the sequence order is $\Lambda_1 \ \Lambda_4 \Lambda_{3,1}\Lambda_{3,2}\Lambda_{3,3} \Lambda_5 \ \Lambda_2$.

\begin{example} If we drop the four bits $b_{18}$, $b_{19}$, $b_{37}$, and $b_{38}$ from the binary message stream $\mathbf{b}$ in Example~\ref{illustrate} and adopt Bridging Scheme~III, the resulting $34$-bit message stream is encoded as $TTGCGTCGC\mathbf{GAGAC}AGCCAACTT\mathbf{CTCTC}$.
\end{example}
\par
\subsection{Probability of Not Detecting Errors} \label{sec_prob}

Now, we study the probability of missing substitution errors when Bridging Schemes~III and II-B are applied. The focus is on detecting errors in codewords, check-sum symbols, and bridging symbols within which message stream bits are encoded. {Below, we start with a simple DNA storage channel where substitution errors dominate the error profile, which allows us to compare the two bridging schemes. 

We assume that we have a $(1-p,p/3,p/3,p/3)-$symmetric DNA storage channel with symbol substitution error rate $p \in [0,1]$. This is a channel with $4$-ary input and $4$-ary output such that 
\begin{align*} &\mathbb{P}(Y = \Lambda | X = \Lambda) = 1-p \textrm{ and } \\
&\mathbb{P}(Y = \Lambda' | X = \Lambda) = p/3, 
\end{align*}
for any letters $\Lambda \in \{A,T,G,C\}$ and $\Lambda' \in \{A,T,G,C\} \setminus \{\Lambda\}$.} For example, Organick et al. performed extensive experiments and reported substitution, deletion, and insertion rates as
$4.5 \times 10^{-3}$, $1.5 \times 10^{-3}$, and $5.4 \times 10^{-4}$, respectively \cite{organick_etal}. We can see that the substitution error rate here is the dominant error rate, and this value of $4.5 \times 10^{-3}$ can be considered a typical value of $p$ based on \cite{organick_etal}.

We start the analysis with Bridging Scheme~III. Suppose  we use codewords in $\mathcal{D}_{3m',\ell}$ and this bridging scheme is applied. For $i \in \{1,2,3\}$, let
\begin{align*} {\mathsf{U}_i} & \triangleq  \textrm{ the event that the check-sum represented by } \\
& \hspace{1.7em} \Lambda_{3,i} \textrm{ is satisfied in the read codeword } \mathbf{c}=\mathbf{c}_1\mathbf{c}_2\mathbf{c}_3 \textrm{,} \\
& \hspace{1.5em} \textrm{ i.e., } \Lambda_{3,i}=\sum^{m'-1}_{k=0} \mathsf{a}_{i,k} \textrm{ }(\mathrm{mod}\,4)\textrm{ for } \mathbf{c}_i \textrm{,} \\
{\mathsf{E}_i} & \triangleq \textrm{ the event that there are errors in } \mathbf{c}_i \textrm{ or in } \Lambda_{3,i} \textrm{,}  \\
P_i& \triangleq \mathbb{P}({\mathsf{U}_i \land \mathsf{E}_i}) \\
&= \textrm{the probability that there are errors in } \mathbf{c}_i \\
& \hspace{1em} \textrm{ or in } \Lambda_{3,i} \textrm{ but are undetected (check-sum satisfied).} 
\end{align*}
Clearly, $P_1=P_2=P_3$ for the three constituent sequences. Next, we give upper bounds for $P_1$.
\par
Observe that if a single error occurs in $\mathbf{c}_1$ or in $\Lambda_{3,1}$, it is detected. Moreover, the probability of no-detection in case $r$ errors occur, where $r \geq 2$, is at most $1/3$ because the number of error patterns at $r$ specific locations which (i) do not affect the check-sum and (ii) do not yield a forbidden pattern is at most $3^{r-1}$ out of $3^r$ possible error patterns. Thus, 
\begin{align} P_1 \leq \frac{1}{3} \left (1-(1-p)^{m'+1}-(m'+1)p(1-p)^{m'} \right ), \nonumber \\ \textrm{ (Bound I).}
\end{align}

In fact, via a more careful analysis, we can give a closer upper bound for $P_1$ as follows. Observe that for $r$ specific locations, the integer-equivalent options of the error patterns that will go undetected are
$$1\text{-}3,2\text{-}2,3\text{-}1 \textrm{ when } r=2, \textrm{ and}$$
$$1\text{-}1\text{-}2,1\text{-}2\text{-}1,2\text{-}1\text{-}1,2\text{-}3\text{-}3,3\text{-}2\text{-}3,3\text{-}3\text{-}2 \textrm{ when } r=3.$$
We denote this number of error-pattern options by $C(r)$. Since every undetected $r$-error pattern comes from a unique $(r-1)$-error pattern that is guaranteed to be detected (its error check-sum is not $0$ $(\mathrm{mod}\,4)$), we have the following non-homogeneous linear recurrence relation
\begin{equation} C(r)=3^{r-1}-C(r-1). \label{recurC} 
\end{equation}
The closed-form solution of this relation that satisfies $C(2)=3$ can be derived using the z-transform, and it is
\begin{align} C(r) &= \frac{3^{r}-3(-1)^{r-1}}{4}. 
\end{align}
Hence, the probability of no-detection in case of $r$ errors, $r \geq 2$, is at most $C(r)/{3^r}=[1-(-1/3)^{r-1}]/4$, and we express a closer upper bound (Bound II) for $P_1$ in (\ref{eqn_bound2}). 
\begin{figure*}
\vspace{-0.3em}
\noindent\makebox[\linewidth]{\rule{\textwidth}{1,5pt}} \\
\begin{align} \label{eqn_bound2} \displaystyle P_1 &\leq  \sum^{m'+1}_{r=2} \bigg[ \frac{1-(-1/3)^{r-1}}{4} \bigg] \binom{m'+1}{r}p^r(1-p)^{m'+1-r}
= \frac{1}{3} \binom{m'+1}{2}p^2(1-p)^{m'-1}+\frac{2}{9} \binom{m'+1}{3}p^3(1-p)^{m'-2} \nonumber \\ &\hspace{+15.0em}+\frac{7}{27} \binom{m'+1}{4}p^4(1-p)^{m'-3}+\dots+\bigg[ \frac{1-(-1/3)^{m'}}{4} \bigg] p^{m'+1}, \textrm{ (Bound II).}
\end{align} \\
\noindent\makebox[\linewidth]{\rule{\textwidth}{1,5pt}}
\end{figure*}
\par
The frame-level probability $P_{\mathrm{un}}$ that there are errors in the codeword $\mathbf{c}$ and/or the symbols $\Lambda_{3,1}\Lambda_{3,2}\Lambda_{3,3}$ but are~undetected is the probability that all the check-sums are satisfied and at least one of the events ${\mathsf{E}_1, \mathsf{E}_2}$, and ${\mathsf{E}_3}$ occurs. Thus, we have the following expression for $P_{\mathrm{un}}$:
\vspace{-0.1em}\begin{align} \label{prob_express} P_{\mathrm{un}} &= \mathbb{P}({\mathsf{U}_1 \land \mathsf{U}_2 \land \mathsf{U}_3 \land (\mathsf{E}_1 \lor \mathsf{E}_2 \lor \mathsf{E}_3)}) \nonumber \\
&= 3 P_1(1-p)^{2(m'+1)}+ 3P_1^2(1-p)^{m'+1} + P_1^3.
\end{align}

As for Bridging Scheme~II-B, the frame-level probability $P_{\mathrm{un}}$ that there are errors in the coded sequence $\mathbf{c} \ \Lambda_4\Lambda_3$, where $\mathbf{c} \in \mathcal{D}_{m,\ell}$, but are undetected can be directly bounded using Bound II in (\ref{eqn_bound2}) by replacing $m'+1$ by $m+2$ (these are $m$ symbols in $\mathbf{c}$, $\Lambda_4$, and $\Lambda_3$).

\begin{figure}
\vspace{-0.5em}
\center
\includegraphics[trim={0.0in 0.0in 0.0in 0.0in}, width=3.2in]{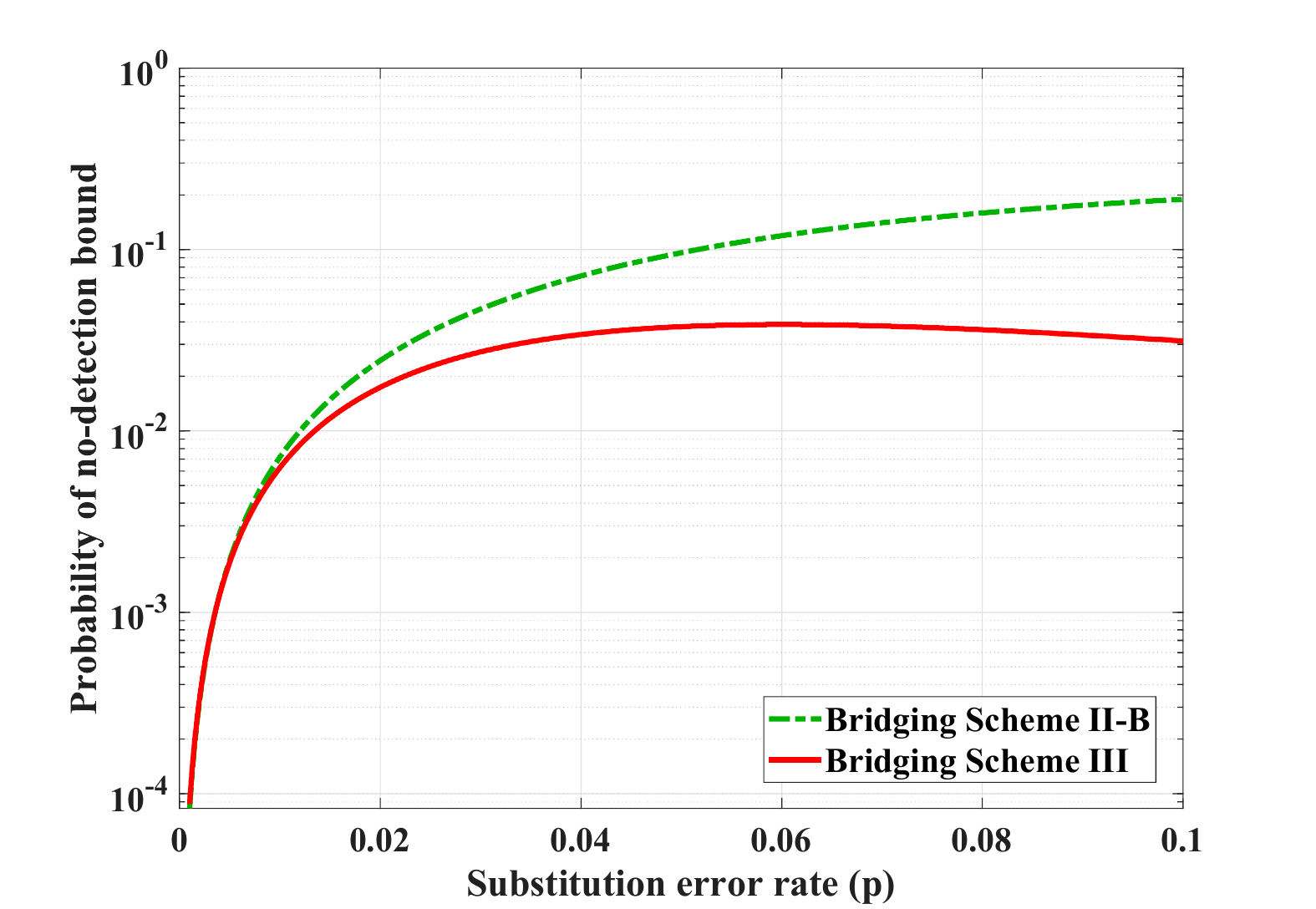}
\vspace{-0.7em}
\caption{{Upper bounds on the probability of no-detection for $\mathcal{D}_{3m',3}$ and Bridging Scheme~III versus $\mathcal{D}_{m,3}$ and Bridging Scheme~II-B, $m'=13$ and $m=21$}.}
\label{fig_detect1}
\vspace{-0.5em}
\end{figure}

\begin{figure}
\vspace{-0.5em}
\center
\includegraphics[trim={0.0in 0.0in 0.0in 0.0in}, width=3.2in]{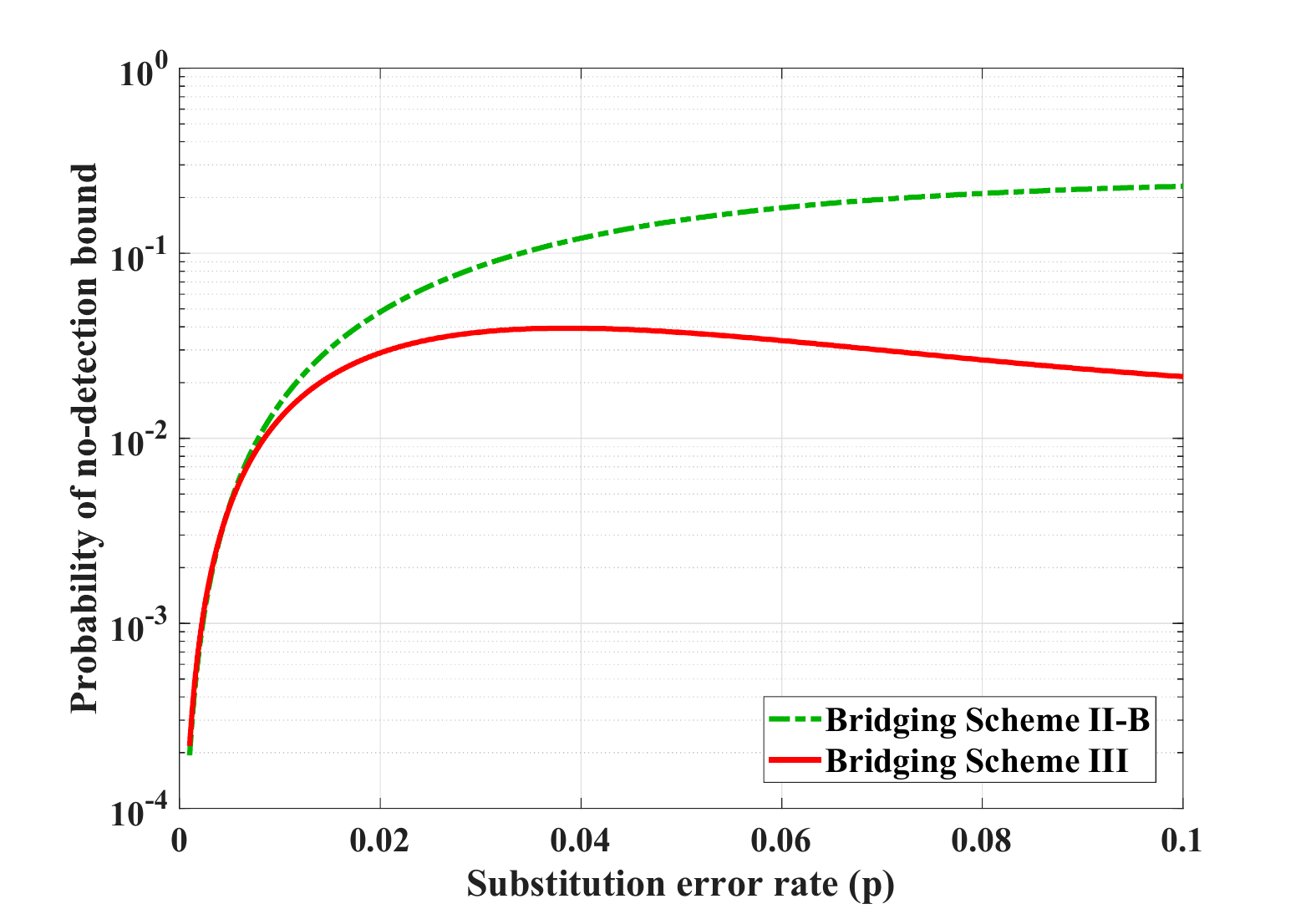}
\vspace{-0.7em}
\caption{{Upper bounds on the probability of no-detection for $\mathcal{D}_{3m',3}$ and Bridging Scheme~III versus $\mathcal{D}_{m,3}$ and Bridging Scheme~II-B, $m'=21$ and $m=33$}.}
\label{fig_detect2}
\vspace{-0.4em}
\end{figure}

Fig.~\ref{fig_detect1} and Fig.~\ref{fig_detect2} compare the upper bound on the probability of no-detection $P_{\mathrm{un}}$ (using Bound~II) of Bridging Scheme~III and Bridging Scheme~II-B at various values of the channel substitution error rate $p$. Fig.~\ref{fig_detect1} compares the no-detection performance of Bridging Scheme~III for the code $\mathcal{D}_{3m',3}$ with $m'=13$ and of Bridging Scheme~II-B for the code $\mathcal{D}_{m,3}$ with $m=21$. The rates of these two codes are $0.8542$ and $0.8636$, respectively. Fig.~\ref{fig_detect2} compares the no-detection performance of Bridging Scheme~III for the code $\mathcal{D}_{3m',3}$ with $m'=21$ and of Bridging Scheme~II-B for the code $\mathcal{D}_{m,3}$ with $m=33$. The rates of these two codes are $0.9028$ and $0.9044$, respectively. The choice of code lengths is such that the rates are close for the two coding schemes being compared. Note that code rate discussions are presented in Section~\ref{sec_rate}, and this rate depends on the D-LOCO code, its bridging, and its balancing penalty.

There are three takeaways from the two figures. First, both bridging schemes offer notably low no-detection probabilities for substitution error rates at or below $0.01$. These error rates are of practical importance in DNA storage systems, and we refer the reader to \cite{organick_etal}, where Illumina NextSeq sequencing is adopted. Second, Bridging Scheme~III outperforms Bridging Scheme~II-B in terms of no-detection probability by up to $1$ order of magnitude in the high substitution error rate regime.\footnote{If the comparison is at fixed frame length for similar rates, this gain becomes even higher since the frame length is $3m'+5$ for Bridging Scheme~III while it is only $m+3$ for Bridging Scheme~II-B.} These error rates are common for sequencing methods such as Oxford Nanopore sequencing \cite{olgica_nano}. Third, the gain of Bridging Scheme~III over Bridging Scheme~II-B increases as the code rate increases, i.e., as the length increases.

{
Observe that if a D-LOCO code is paired with an error-correction code such that the former is the inner code (closer to the channel), achieving low no-detection probability makes the task of the error-correction decoder easier.} Observe also that Bridging Scheme~III, because of the higher code length required to achieve the same rate, incurs higher storage overhead and error propagation compared with Bridging Scheme~II-B (see the following sections, \cite{ahh_ps}, and \cite{ahh_general}). Finally, we note that similar conclusions can also be reached for $\ell > 3$.

%\begin{align*} P_{\mathrm{un}}=&\mathbb{P}(u_1 \land u_2 \land u_3 \land (e_1 \lor e_2 \lor e_3)) \\
%=&  P_1(1-p)^{m'+1}(1-p)^{m'+3} + (1-p)^{m'+1}P_2(1-p)^{m'+3} + (1-p)^{m'+1}(1-p)^{m'+1}P_3 \\
%&+ P_1P_2(1-p)^{m'+3} + P_1(1-p)^{m'+1}P_3+(1-p)^{m'+1}P_2P_3 + P_1P_2P_3 \\
%=& P_1^2((1-p)^{m'+3}+P_3)+P_1(1-p)^{m'+1}((1-p)^{m'+3}+P_3) \\
%+& P_1(1-p)^{m'+1}((1-p)^{m'+3}+P_3)+(1-p)^{2(m'+2)}((1-p)^{m'+3}+P_3) -  (1-p)^{m'+1}(1-p)^{m'+1}(1-p)^{m'+3} \\
%=& (P_1+(1-p)^{m'+1})^2((1-p)^{m'+3}+P_3))- (1-p)^{m'+1}(1-p)^{m'+1}(1-p)^{m'+3} \\
%=&  \frac{1}{9}\bigg(1+2(1-p)^{m'+1}-(m'+1)p(1-p)^{m'}\bigg)^2\frac{1}{3}\bigg(1+2(1-p)^{m'+3}-(m'+3)p(1-p)^{m'+2}\bigg)- (1-p)^{3m'+5} \\
%=&  \frac{1}{27}\bigg(1+(1-p)^{m'}(2-(m'+3)p)\bigg)^2\bigg(1+(1-p)^{m'+2}(2-(m'+5)p)\bigg)- (1-p)^{3m'+5}
%\end{align*}

%%%%%%%%%%%%%%%%%%%%%%%%%%%%%%%%%%%%%
\section{Algorithms and Balancing the DNA Sequence} \label{sec_balance}

{In this section, we first present the encoding and decoding algorithms of D-LOCO codes equipped with Bridging Scheme~I, and then discuss how to balance the $GC$-content of the designed DNA sequence.
}
\begin{algorithm} 
{
\caption{{Encoding D-LOCO Codes with Bridging Scheme~I}}
\begin{algorithmic}[1]
\State \textbf{Inputs:} Incoming stream of binary messages and the highest run-length allowed $\ell$.
\State Use the cardinalities $N_{\mathrm{D}}(r, \ell)$, $r \in \{\ell+1, \ell+2, \dots\}$ (computed offline by (\ref{card1})), where $N_{\mathrm{D}}(r, \ell) = 4^r$ for $0 < r \leq \ell$ and $N_{\mathrm{D}}(0,\ell)=4/3$.
\State Specify $m$ as the smallest $r$ in Step~2 to achieve the desired rate. Then, $s = \lfloor \log_2 N_{\mathrm{D}}(m, \ell) \rfloor$.
\State \textbf{for} each incoming message $\bold{b}$ of length $s+1$ \textbf{do}
\State \hspace{2ex} Compute $g(\bold{c})=\mathrm{decimal}(\bold{b})$.
\State \hspace{2ex} Initialize $\mathrm{residual}$ with $g(\bold{c})$ and $c_i$ with $\zeta$ for $i \geq m$. \textit{($\zeta$ indicates out of codeword bounds)}
\State \hspace{2ex} \textbf{for} $i \in \{m-1, m-2, \dots, 0\}$ \textbf{do} \textit{(in order)}
\State \hspace{4ex} For each $\chi \in \{T,G,C\}$, set $c_i = \chi$ (temporarily), determine the coefficients $\mathsf{a}_{i,k},\mathsf{t}_{i,k}, \mathsf{g}_{i,k}$ for $1 \leq k \leq \ell$ based on (\ref{coeffs_gen}), and then compute $M(i,j, \chi) = \sum^{j}_{k=1} (\mathsf{a}_{i,k}+\mathsf{t}_{i,k}+\mathsf{g}_{i,k}) $ for all $1 \leq j \leq  \ell$. 
\State \hspace{4ex} \textbf{for} $\chi \in \{T,G,C\}$ \textbf{do}
\State \hspace{6ex} Compute $\mathrm{contrib}(i, \chi) = \frac{3}{4}\sum^{\ell}_{j=1} M(i,j,\chi) N(i+j-\ell)$.
\State \hspace{4ex} \textbf{end for}
\State \hspace{4ex} \textbf{if} $\mathrm{residual} \geq \mathrm{contrib}(i,C)$ \textbf{then}
\State \hspace{6ex} Encode $c_i = C$.
\State \hspace{4ex} \textbf{else if} $\mathrm{residual} \geq \mathrm{contrib}(i,G)$ \textbf{then}
\State \hspace{6ex} Encode $c_i = G$.
\State \hspace{4ex} \textbf{else if} $\mathrm{residual} \geq \mathrm{contrib}(i,T)$ \textbf{then}
\State \hspace{6ex} Encode $c_i = T$.
\State \hspace{4ex} \textbf{else}
\State \hspace{6ex} Encode $c_i = A$.
\State \hspace{4ex} \textbf{end if}
\State \hspace{4ex} $\mathrm{residual} \leftarrow \mathrm{residual} - \mathrm{contrib}(i,c_i)$, $c_i \neq A$.
\State \hspace{4ex} \textbf{if} (not first codeword) $\land$ ($i = m-1$) \textbf{then}
\State \hspace{6ex} Bridge with a symbol other than $c_{m-1}$ or the right-most symbol of the previous codeword based on the last bit of the incoming message as described in Subsection~\ref{bridging_scheme_1}.
\State \hspace{4ex} \textbf{end if}
\State \hspace{2ex} \textbf{end for}
\State \textbf{end for}
\State \textbf{Output:} Outgoing stream of bridged D-LOCO codewords.
\end{algorithmic}
\label{DLOCO_enc}
}
\end{algorithm}

\begin{algorithm}
{
\caption{{Decoding D-LOCO Codes with Bridging Scheme~I}} 
\begin{algorithmic}[1]
\State \textbf{Inputs:} Incoming stream of $4$-ary D-LOCO codewords, in addition to $m$ and $s$. 
\State Use the cardinalities $N_{\mathrm{D}}(r, \ell)$, $r \in \{\ell+1, \ell+2, \dots, m-1\}$ (computed offline by (\ref{card1})), where $N_{\mathrm{D}}(r, \ell) = 4^r$ for $0 < r \leq \ell$ and $N_{\mathrm{D}}(0,\ell)=4/3$.
\State \textbf{for} each incoming codeword $\bold{c}$ of length $m$ \textbf{do}
\State \hspace{2ex} Initialize $g(\bold{c})$ with $0$ and $c_i$ with $\zeta$ for $i \geq m$. \textit{($\zeta$ indicates out of codeword bounds)}
\State \hspace{2ex} \textbf{for} $i \in \{m-1, m-2, \dots, 0\}$ \textbf{do} \textit{(in order)}
\State \hspace{4ex} Determine the coefficients $\mathsf{a}_{i,k},\mathsf{t}_{i,k}, \mathsf{g}_{i,k}$ for $1 \leq k \leq \ell$ based on (\ref{coeffs_gen}). 
\State \hspace{4ex} Compute  $\mathrm{contrib}(c_i)=\frac{3}{4} \sum^{\ell}_{j=1} \sum^{j}_{k=1} (\mathsf{a}_{i,k}+\mathsf{t}_{i,k}+\mathsf{g}_{i,k}) N(i+j-\ell)$.
\State \hspace{4ex} $g(\bold{c}) \leftarrow g(\bold{c}) + \mathrm{contrib}(c_i)$.
\State \hspace{2ex} \textbf{end for}
\State \hspace{2ex} Compute $\bold{b}=\mathrm{binary}(g(\bold{c}))$, which has length $s$.
\State \hspace{2ex} Concatenate the bit encoded in the next bridging symbol (as described in Subsection~\ref{bridging_scheme_1}). Then, skip this bridging symbol.
\State \textbf{end for}
\State \textbf{Output:} Outgoing stream of binary messages.
\end{algorithmic}
\label{DLOCO_dec}
}
\end{algorithm}

We now discuss how to balance the DNA sequence or stream of bridged D-LOCO codewords in a way that incurs minimal rate penalty. {Balancing is needed to have a $GC$-content close to $\% 50$ for the DNA strand in order to achieve high reliability of the strand synthesized \cite{ross_etal, schwartz_etal}.
} 
\begin{definition} \label{dispar} The disparity, denoted by $p(\mathbf{c}$), of a codeword $\mathbf{c}$ in $\mathcal{D}_{m,\ell}$ is defined as the difference between the total number of symbols $G,C$ and the total number of symbols $A,T$, i.e.,
\begin{equation}
p(\mathbf{c})=|G|+|C|-|A|-|T|,
\end{equation}
where $|\Lambda|$ denotes the number of occurrences of the symbol $\Lambda$ in $\mathbf{c}$. The absolute value of $p(\mathbf{c})$ is called the absolute disparity of $\mathbf{c}$. The global disparity of a DNA sequence/stream of length $M$ consisting of $K$ codewords $\mathbf{c}_i$, $i$ in $\{1, 2, \dots, K\}$, along with their bridging symbols is defined as the absolute value of the sum of disparities of all codewords and their bridging in the DNA sequence, and it is given by
\vspace{-0.1em}\begin{equation}
p^\mathrm{global} = \bigg|  \sum\limits^{K}_{i=1}  p(\mathbf{c}_i) + \psi \bigg| \in [0,M],
\end{equation}
where $\psi$ is the cumulative disparity resulting from all bridging symbols.\footnote{For ranges of integer variables, such as the disparity, the notation $[a,b]$ means $\{a, a+1, \dots, b\}$ and the notation $(a,b)$ means $\{a+1, a+2, \dots, b-1\}$. We use $[a, b]$ and $(a, b)$ for brevity.} The global disparity fraction is defined as $p^\mathrm{global}/M \in [0,1]$. Note that if the global disparity fraction is in $[0,0.1]$, then the $GC$-content of the DNA sequence is between $\% 45$ and $\%55$ (see also \cite{song_cai_etal} for further discussions regarding the $GC$-content of a DNA sequence).
\end{definition}

Observe that our bridging schemes introduced in Section~\ref{sec_bridge} contribute to ensuring low global disparity fraction of the DNA sequence. For some of our bridging schemes, the global disparity fraction is in $[0,\frac{1}{K})$ as discussed in Remark~\ref{rem_dispar}.

\subsection{Balancing the D-LOCO Codes of Odd Lengths} 

Suppose each message is encoded into a codeword in $\mathcal{D}_{m,\ell}$, where $m$ is odd and Bridging Scheme~I is applied. Note that each sequence of odd length has necessarily nonzero disparity. For a codeword $\mathbf{c} \in \mathcal{D}_{m,\ell}$, let $\mathbf{\bar{c}}$ be the codeword in $\mathcal{D}_{m,\ell}$ obtained by replacing each $A, T , G, C$'s in $\mathbf{c}$ by $C, G, T, A$, respectively. Here, $\mathbf{\bar{c}}$ will be called the {\textit{complement}} of $\mathbf{c}$ as $g(\mathbf{\bar{c}})+g(\mathbf{c}) = N(m)-1$ (see also \cite{ahh_loco}). For example, the complement of $GAATC$ is $TCCGA$ and vice versa. Observe that this is possible for any $\mathbf{c} \in \mathcal{D}_{m,\ell}$ because of the intrinsic symmetry of the D-LOCO code stemming from the intrinsic symmetry of the set of forbidden patterns. Observe also the relation between disparities: $p(\mathbf{\bar{c}})=-p(\mathbf{c})$.

\begin{lemma} \label{lemma_dispar} Using the above setup, there is a balancing procedure so that a D-LOCO coded sequence of length $K(m+1)$ has global disparity in $[-m-1,m+1]$, for odd $m$. In particular, the order of the global disparity is $O(m)$, and thus, it is independent of the overall sequence length for fixed $m$.
\end{lemma}
\begin{IEEEproof} We prove this by induction on $K$. If $K=1$, the result is immediate. For an induction hypothesis, assume that a coded sequence $\mathbf{s}$ of length $K(m+1)$ has global disparity in $[-m-1,m+1]$. {The $(K+1)^{\textrm{th}}$ codeword $\mathbf{c}$ can be replaced, if necessary, by its complement $\mathbf{\bar{c}}$ so that it has opposite disparity sign to the one of $\mathbf{s}$ (while the new bridging letter can have any disparity).} Thus, the resulting coded sequence will have disparity in $[-m-1,m+1]$. The minimum (resp., maximum) disparity is achieved, for example, if $\mathbf{s}$ has disparity $-m-1$ (resp., $m+1$) and the new codeword along with its bridging symbol have disparity $0$. This finishes the induction argument.
\end{IEEEproof}

\begin{table*}
\vspace{-0.2em}
\caption{Normalized Rates When Different Bridging Schemes Are Applied at Various Lengths for $\ell=3$}
\vspace{-0.1em}
\centering
\scalebox{1.1}
{
\begin{tabular}{|c|c|c|c|c|c|c|c|}
\hline
\makecell{$m$} & \makecell{$R_1$} & \makecell{$m$} & \makecell{$R_2$} & \makecell{$m$} & \makecell{$R_3$}  & \makecell{$m'$} & \makecell{$R_4$}
\\
\hline
$9$ & $0.8500$ & $9$ & $0.7500$ & $9$ & $0.7083$ & $5$ & $0.7000$ \\
\hline
$13$ & $0.8929$ & $13$ & $0.8125$ & $13$ & $0.7812$ & $7$ & $0.7692$ \\
\hline
$21$ & $0.9318$ & $21$ & $0.8750$ & $21$ & $0.8541$ & $11$ & $0.8421$\\
\hline
$33$ & $0.9559$ & $33$ & $0.9167$ & $33$ & $0.9027$  & $17$ & $0.8929$ \\
\hline
$51$ & $0.9712$  & $51$ & $0.9444$ & $51$ & $0.9351$ & $21$ & $0.9044$ \\
\hline
$99$ & $0.9800$ & $99$ & $0.9657$ & $99$ & $0.9607$ & $33$ & $0.9375$ \\
\hline
Capacity & $0.9912$ & Capacity & $0.9912$ & Capacity & $0.9912$  & Capacity & $0.9912$ \\
\hline
\end{tabular}}
\label{table1}
\end{table*}

\begin{remark} \label{rem_balance} Using the above balancing procedure, there are two codewords to encode each message. Thus, one can use at most ``half of the D-LOCO codebook'' for distinct messages, i.e., precisely $2^{\lfloor \log_2(\frac{1}{2}N(m)) \rfloor}$-many codewords correspond to unique messages each. Hence, our D-LOCO codes achieve the minimum possible rate loss; specifically, the one-bit penalty in the message length achieved by our balancing procedure results in the minimum rate loss. Moreover, this balancing penalty does not affect the capacity achievability of D-LOCO codes (see Section VI in \cite{ahh_loco} for a thorough discussion).
\end{remark}

\begin{remark} \label{rem_dispar} We can similarly show the following: \\
(i) For odd $m$, a coded sequence of length $K(m+3)$, where Bridging Scheme~II-A is applied, has global disparity in $[-m-2K-1,m+2K+1]$,  \\
(ii) If we go without the extra $1$-bit of information encoded in the bridging symbol $\Lambda_5$ and adopt Bridging Scheme~II-B, we can attain global disparity in the range $[-m-1,m+1]$ for a coded sequence of length $K(m+3)$. In this case, the global disparity fraction is in the range $[0,\frac{1}{K})$ and for $K=10$, the $GC$-content of the DNA sequence is guaranteed to be between $45\%$ and $55\%$. An even better result is obtained for $K=25$, where the $GC$-content of the DNA sequence is guaranteed to be between $48\%$ and $52\%$. \\
(iii) Similarly and for odd $m=3m'$, a coded sequence of length $K(m+5)$ has global disparity in $[-m-1,m+1]$ when Bridging Scheme~III is applied.
\end{remark}

Observe that our disparity range is for any number of codewords $K$, small, medium, or large. Therefore, our balancing procedure not only achieves a global balancing criterion, but also achieves a local balancing one.

%
%\subsection{Balancing the even-length codes} 
%Suppose each message is encoded into a codeword in $\mathcal{D}_{m,\ell}$ where $m$ is even and a single letter is used for bridging. Note that a codeword in $S_{m,l}$ can have any disparity in the range $[-m,m]$.
%\claim Under the above assumption, a coded sequence of length $K(m+1)$ has global disparity in $[\min\{-m-1,-K+1\}-1,\max\{m+1,K-1\}+1]$.
%\proof We prove this by induction on $K$. If $K=1$, it is immediate. For an induction hypothesis, assume that a coded sequence $\mathbf{s}$ of length $K(m+1)$ has global disparity in $[\min\{-m-1,-K+1\}-1,\max\{m+1,K-1\}+1]$. If $\mathbf{m}$ has zero disparity, then for any concatenated codeword, the resulting coded sequence will have disparity in $[-m-1,m+1]$. If $\mathbf{m}$ has a negative disparity, then by our balancing procedure, the concatenated codeword can be chosen in a way that it has nonnegative disparity (while the bridging letter might have any disparity) so that the resulting coded sequence will have global disparity in  $[\min\{-m-1,-K+1\},m]$.  Similarly, if $\mathbf{m}$ has a positive disparity, then by our balancing procedure, the concatenated codeword can be chosen in a way that it has nonpositive disparity so that the resulting coded sequence will have global disparity in  $[-m,\max\{m+1,K-1\}]$. This finishes the induction argument.

%%%%%%%%%%%%%%%%%%%%%%%%%%%%%%%%%%%%%
\section{Achievable Rates for $\ell=3$ and Literature Comparison} \label{sec_rate}

{An \textit{FSTD} is a state diagram that represents the infinitude of a sequence in which, some chosen patterns are forbidden. The FSTD corresponding to the set $\mathcal{F}$ of patterns consisting of runs of length $\ell+1$ (see Definition~\ref{D-LOCO_dna}) is given in Fig.~\ref{fig:FSTD}.} This FSTD has $\ell$ states, and state $F_j$ represents the case where the last $j+1$ generated symbols are $\Lambda'\mathbf{\Lambda}^{j}$, where $\Lambda \in \{A,T,G,C\}$ and $\Lambda' \in \{A,T,G,C\} \setminus \{\Lambda\}$ arbitrarily. In particular, for $\ell=3$, state $F_1$ represents the case where the last two generated symbols are $\Lambda'\Lambda$, $F_2$ represents the case where the last three symbols are $\Lambda'\Lambda \Lambda$, and $F_3$ represents the case where the last four symbols are $\Lambda'\Lambda\Lambda\Lambda$, where $\Lambda \in \{A,T,G,C\}$ and $\Lambda' \in \{A,T,G,C\} \setminus \{\Lambda\}$ arbitrarily. The symbols on the directed transition edges represent the currently generated symbols. The adjacency matrix of the FSTD for fixed $\ell \geq 3$ is denoted by $\mathbf{F}^{(\ell)}$, and it is of size $\ell \times \ell$. The entry $f_{h,p}$, $1 \leq h,p \leq \ell$, is the number of times state $F_h$ is connected to state $F_p$ (from $F_h$ to $F_p$). Consequently, 
\begin{equation} \mathbf{F}^{(\ell)} = \left[ \begin{array}{c | c c c} 3 &  &  & \\  \vdots &  & \mathbf{I}_{\ell-1} & \\ 3 & & & \\ \hline 3 & 0 & \cdots & 0 \end{array} \right], 
\end{equation}
where $\mathbf{I}_{\ell-1}$ is the identity matrix of size $(\ell-1) \times (\ell-1)$, and
\begin{equation} \mathbf{F}^{(3)} = \begin{bmatrix} 3 & 1 & 0 \\ 3 & 0 & 1 \\ 3 & 0 & 0 \end{bmatrix}. 
\end{equation}
The characteristic polynomial of $\mathbf{F}^{(3)}$ is then: 
\be \mathrm{det}(\beta I-\mathbf{F}^{(3)}) = \beta^3-3\beta^2-3\beta-3. \ee
Therefore, the largest real positive eigenvalue of $\mathbf{F}^{(3)}$ is $\beta_{\mathrm{max}}=3.9514$, and the normalized capacity for $\ell=3$ is $C^{(3)}=\log_4(\beta_{\mathrm{max}})=0.9912$.

\tikzset{node distance=2cm,
every state/.style={circle, draw, semithick,minimum size=2.4em},
initial text={},
double distance=2pt,
every edge/.style={ draw,
->,>=stealth', auto, semithick}}
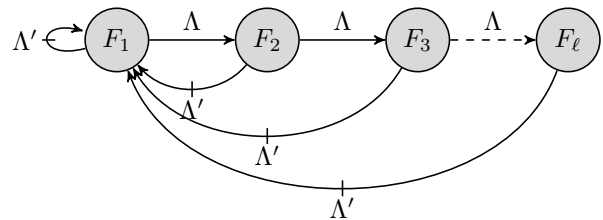
\begin{figure}
\vspace{-0.5em}
\centering
\begin{tikzpicture}
\node[state,  fill=gray!30] (A) {$F_1$};
\node[state, fill=gray!30, right of=A] (B) {$F_2$}; 
\node[state, fill=gray!30, right of=B] (C) {$F_3$};
\node[state, fill=gray!30, right of=C] (D) {$F_{\ell}$};
\draw (A) edge node[above] {$\Lambda$} (B);
\draw (B) edge node[above] {$\Lambda$} (C);
\draw (A) edge [loop left=60, strike through] node {$\Lambda'$} (A);
\draw (B) edge [bend left=50, strike through] node[below] {$\Lambda'$} (A);
\draw (C) edge [bend left=60, strike through] node[below] {$\Lambda'$} (A);
\draw (D) edge [bend left=70, strike through] node[below] {$\Lambda'$} (A);
\draw (C) edge[dashed] node[above] {$\Lambda$} (D);
\end{tikzpicture}
\caption{An FSTD of $\mathcal{F}$-constrained sequences, where $\Lambda' \in \{A,T,G,C\} \setminus \{\Lambda\}$. Note that $\Lambda$ represents the last generated symbol at any state. Upon entering $F_1$, $\Lambda$ is always updated to become the relevant $\Lambda'$, and $\Lambda'$ symbols on different FSTD edges are not necessarily the same.}
\label{fig:FSTD}
\vspace{-0.5em}
\end{figure}

\begin{table*}
\caption{Comparison Between Our Coding Schemes Eliminating Homopolymers of Length $\geq 4$ With Increasing Rates}
\vspace{-0.1em}
\centering
\scalebox{1.1}
{
\begin{tabular}{|c|c|c|c|c|c|}
\hline
 \multicolumn{1}{|c|}{\makecell{Strand sequence \\ length}} & \multicolumn{1}{|c|}{\makecell{$GC$-content}} & \multicolumn{1}{|c|}{\makecell{Single substitution \\ error detection}}  &\multicolumn{1}{|c|}{\makecell{Normalized \\ rate}} & \multicolumn{1}{|c|}{\makecell{Storage overhead}} & \multicolumn{1}{|c|}{\makecell{Length and \\ bridging method}}  \\
\hline
 288 & $45\%-55\%$ & \textrm{Yes} & $0.8125$ & $703$ bits & $m'=9, K=9$, III\\
\hline
240 & $45\%-55\%$ & \textrm{Yes} & $0.8541$ & $421$ bits & $m=21, K=10$, II-B \\
\hline
180 & $45\%-55\%$ & \textrm{No} & $0.9167$ & $273$ bits & $m=17, K=10$, I \\
\hline
324 & $42\%-58\%$ & \textrm{Yes} & $0.9167$ & $1057$ bits  & $m=33, K=9$, II-A \\
\hline
220& $45\%-55\%$ & \textrm{No} & $0.9318$ & $421$ bits  & $m=21, K=10$, I \ \\
\hline
\end{tabular}}
\label{table2}
\end{table*}
\vspace{+0.3em}

\par We now discuss the finite-length rates of D-LOCO codes for each bridging scheme we propose and with the requirement of balancing. 

\begin{itemize} 
\item The normalized rate $R_1$ of the D-LOCO code $\mathcal{D}_{m,\ell}$ when Bridging Scheme~I is adopted is given by
\begin{equation}  \label{R1} R_1=\frac{\lfloor \log_2(\frac{1}{2}N(m))\rfloor+1}{2(m+1)}=\frac{\lfloor \log_2(N(m)) \rfloor}{2(m+1)}. 
\end{equation}

\item The normalized rate $R_2$ of the D-LOCO code $\mathcal{D}_{m,\ell}$ when Bridging Scheme~II-A is adopted is given by
\begin{equation} \label{R2} R_2=\frac{\lfloor \log_2(\frac{1}{2}N(m))\rfloor+2}{2(m+3)}=\frac{\lfloor \log_2(N(m)) \rfloor +1}{2(m+3)}.
\end{equation}

\item The normalized rate $R_3$ of the D-LOCO code $\mathcal{D}_{m,\ell}$ when Bridging Scheme~II-B is adopted is given by
\begin{equation} \label{R2} R_3=\frac{\lfloor \log_2(\frac{1}{2}N(m))\rfloor+1}{2(m+3)}=\frac{\lfloor \log_2(N(m)) \rfloor}{2(m+3)}.
\end{equation}

\item The normalized rate $R_4$ of the D-LOCO code $\mathcal{D}_{3m',\ell}$ when Bridging Scheme~III is adopted is given by
\begin{equation} \label{R4} R_4=\frac{\lfloor \log_2(\frac{1}{2}N(3m'))\rfloor}{2(3m'+5)}=\frac{\lfloor \log_2(N(3m')) \rfloor -1}{2(3m'+5)}.
\end{equation}
\end{itemize}

For balanced D-LOCO codes, Table~\ref{table1} shows the normalized rates at various values of $m$ when different bridging schemes are applied using (\ref{R1})--(\ref{R4}).

Next, we present brief comparisons between D-LOCO codes and other codes designed for similar goals: 
\begin{enumerate}
\item Nguyen et al. \cite{nguyen_etal} introduced constrained codes that are capable of correcting a single error of any type based on enumerative approach (Method~A) as well as a sequence replacement method (Method~B) for DNA data storage. While their codes from Method~A are rate-wise efficient, D-LOCO codes offer higher rates at lower lengths and also simpler encoding-decoding compared with their unrank-rank approach (see also \cite{ahh_asym}).

\item Improving \cite{song_cai_etal}, Wang et al. \cite{wang_etal} offered an efficient coding scheme, a product of which is a code with a normalized rate of $0.8125$ at length $8$ that eliminates homopolymers of length at least $4$ and has a guaranteed $40\%-60\%$ $GC$-content. This code is based on look-up tables. The nature of our work differs from theirs mainly in three aspects. First, we have more control over the disparity, or equivalently the $GC$-content, of the DNA sequence, both globally and locally. One can choose a practical code length and bridging scheme to achieve low global disparity fraction such as $0.1$, equivalently $45\%-55\%$ $GC$-content (see Table~\ref{table2}), or even lower. Second, our encoding and decoding algorithms are based on small-size adders, offering low complexity. Third, we also propose novel ideas for bridging to guarantee a single substitution error detection.

\item Park et al. \cite{park_lee_no} proposed an iterative encoding algorithm that has a mapping table with $48$ $3$-tuple $4$-ary entries as a building block for a constrained code that addresses the $GC$-content and the maximum homopolymer length requirements. There is a similar table for the iterative decoding algorithm as well. For instance, they achieve a normalized code rate of $0.9165$ and $45\%-55\%$ $GC$-content range. Our approach differs from theirs in terms of encoding-decoding, and we offer local balancing as well as substitution error detection. Moreover, the storage overhead associated with their encoding-decoding algorithms is higher than that associated with our algorithms since we only need to store the cardinalities to be used in the execution of the rule (see Section~\ref{sec_props}).

{ 
\item Liu et al. in \cite{liu_he_tang} proposed a constrained coding scheme based on the unrank-rank approach, enumerating all constrained sequences that achieve a $GC$-content in $[0.5 - \epsilon, 0.5+\epsilon]$ and do not contain homopolymers of length larger than $\ell$. Their encoding-decoding algorithms have polynomial execution time and storage overhead. They also offer a coding scheme that satisfies a local $GC$-content constraint on sequence prefixes in order to further improve immunity against errors. Since both are constructed by employing enumerative coding techniques, their codes and our D-LOCO codes are capacity-achieving. Since the time complexity of the encoding-decoding algorithms of our D-LOCO codes is algorithmically $O(m)$ and implementation-wise $O(m^2)$ with respect to the code length $m$ (see the detailed discussion at the end of Subsection~\ref{complexity}), they are at least as efficient as those of the codes proposed in \cite{liu_he_tang}. D-LOCO codes also have only $O(m^2)$ storage overhead (see (\ref{storage_complexity})). We opt to use moderate lengths in order to mitigate codeword-to-message error propagation. We also take finite-length features into account, such as the effect of bridging and flooring (see (\ref{R1})--(\ref{R4})), while computing our rates in Table~\ref{table1}. Furthermore, our codes offer substitution error-detection, reconfigurability and efficient local balancing (see Subsection~\ref{local_balancing}).
} 
\end{enumerate} 

%%%%%%%%%%%%%%%%%%%%%%%%%%%%%%%%%%%%%
\section{Some Properties of D-LOCO codes}\label{sec_props}

We now discuss additional properties of D-LOCO codes regarding 
{\begin{itemize} 
\item[--] capacity-achievability, 
\item[--] complexity and storage, 
\item[--] ease of reconfigurability, 
\item[--] error propagation,
\item[--] parallelism, and
\item[--] local $GC$-content balance.
\end{itemize}
}
\subsection{Capacity-Achievability} \label{capacity_achievability}

The {\textit{normalized asymptotic rate}}, i.e., the normalized capacity, of a constrained code having length $m$ and forbidding the patterns in $\mathcal{F}$ is 
\begin{equation} C^{(\ell)}=\underset{m \rightarrow \infty}{\lim} \frac{\log_2(N(m)) }{2m}, 
\end{equation} 
which is the base-$4$ logarithm of the largest real positive root of the polynomial $$x^{\ell}-3x^{\ell-1}-3x^{\ell-2}-\dots-3$$ (consistent with the recursive relation in Proposition \ref{prop_card}) \cite{shan_const}.

We note the following about D-LOCO codes: \\
(i) All codewords satisfying the constraint are included in the D-LOCO codebook. \\
(ii) The number of added bits/symbols for bridging is independent of $m$ for all bridging schemes proposed. \\
(iii) The rate loss due to our balancing has the highest possible vanishing rate with respect to $m$, which is $O(1)/O(m)$. This rate loss asymptotically goes to $0$.

Using (i) and (ii) along with all the rate equations (\ref{R1})--(\ref{R4}) for all bridging schemes, we deduce that a D-LOCO code $\mathcal{D}_{m,\ell}$ is \textbf{capacity-achieving}. Using (iii) and the same set of rate equations, we deduce that our balancing procedure has no effect on capacity-achievability.

\subsection{Complexity and Storage Overhead} \label{complexity}

The encoding algorithm of D-LOCO codes is based on comparisons and subtractions, whereas the decoding algorithm mainly uses additions. The size of the adders used to perform these tasks is the base-$2$ logarithm of the maximum value $g(\mathbf{c})$ can
take that corresponds to a message, and it is given by:
$$\left \lfloor \mathrm{log}_2 \left ( \frac{1}{2} N(m) \right ) \right \rfloor,$$
which is the message length (see Remark~\ref{rem_balance}). Table~\ref{table3} illustrates the encoding and decoding complexity of D-LOCO codes through the size of the adders to be used for different lengths and different bridging schemes adopted. For example, for a D-LOCO code with Bridging Scheme~II-A, if a rate of $0.75$ is satisfactory, adders of size just $16$ bits are all that is needed. In case the rate needs to be $0.8750$, adders of size $40$ bits should be used. Moreover, for a D-LOCO code with Bridging Scheme~III, if a rate of $0.70$ is satisfactory, small adders of size just $28$ bits are all that is needed. In case the rate needs to be about $0.8421$, adders of size $64$ bits should be used. All rates are normalized.

\begin{table}
\caption{Adder Sizes Required at Various Lengths for $\ell=3$ When Different Bridging Schemes Are Applied}
\vspace{-0.1em}
\centering
\scalebox{1.1}
{
\begin{tabular}{|c|c|c|c|c|c|c|c|}
\hline
\multicolumn{2}{|c|}{\makecell{I, II-A, or II-B is applied}} & \multicolumn{2}{|c|}{\makecell{III is applied}}  \\
\hline
$m$ & \textrm{Adder size}  & $m'$ & \textrm{Adder size} \\
\hline
$9$ & $16 \textrm{ bits}$  & $5$ & $28 \textrm{ bits}$\\
\hline
$21$ & $40 \textrm{ bits}$  & $11$ & $64 \textrm{ bits}$\\
\hline
$33$ & $64 \textrm{ bits}$ & $17$ & $100 \textrm{ bits}$ \\
\hline
$51$ & $100 \textrm{ bits} $  & $21$ & $123 \textrm{ bits} $ \\
\hline
$99$ & $195 \textrm{ bits} $ & $33$ & $195 \textrm{ bits} $ \\
\hline
\end{tabular}}
\label{table3}
\end{table}
\vspace{+0.3em}

As for the storage overhead, we need to store the values in $\{3N(i)/4 \textrm{ } | \textrm{ } 0 \leq i \leq m-1\}$, for code length $m$, offline in order to execute the encoding-decoding rule in (\ref{eqn_rule3}) and also in (\ref{eqn_rulegen}). Thus, this overhead (in bits) is expressed as follows:
\begin{equation} \label{storage_complexity}
\mathrm{storage} = \sum_{i=0}^{m-1} \left \lceil \mathrm{log}_2 \left ( \frac{3}{4} N(i) \right ) \right \rceil.
\end{equation}
For say $m=27$ (and $m'=9$), we need to store the values in $\{3N(i)/4 \textrm{ } | \textrm{ } 0 \leq i \leq 26\}$ for encoding-decoding. This requires only $703$-bit offline memory, where we achieve a normalized rate of $0.8125$ by adopting Bridging Scheme~III. The same rate is achieved in \cite{wang_etal} for a content-balanced RLL code at length $8$ and binary message length $13$, but this technique is based on lookup tables, which incur higher complexity. {Observe that all constrained coding techniques that are based on lookup tables incur higher storage overhead.}
\par 
\indent For $m=17$, we need to store the values in $\{3N(i)/4 \textrm{ } | \textrm{ } 0 \leq i \leq 16\}$ for encoding-decoding. This requires $273$-bit offline memory, where we achieve a normalized rate of $0.9167$ and global disparity fraction $0.1$ (equivalently, $45\%-55\%$ $GC$-content) by adopting Bridging Scheme~I for $K=10$ (see Lemma~\ref{lemma_dispar}).
Alternatively, for $m=33$, we need to store the values in $\{3N(i)/4 \textrm{ } | \textrm{ } 0 \leq i \leq 32\}$ for encoding-decoding. This requires $1057$-bit offline memory, where we achieve a normalized rate of $0.9167$ and global disparity fraction 
$$\frac{p^{\mathrm{global}}}{M}=\frac{33+2 \times 9+1}{9(33+3)}=0.1605$$
(equivalently, $42\%-58\%$ $GC$-content) by adopting Bridging Scheme~II-A for $K=9$ (see Remark~\ref{rem_dispar} in Section~\ref{sec_balance} above). A rate of $0.9165$ along with $45\%-55\%$ $GC$-content range are achieved in \cite{park_lee_no}, and their iterative encoding-decoding algorithms are based on $48$-entry mapping tables.
\par
Observe that if the DNA storage system can afford adders of higher sizes, our D-LOCO codes can offer notably higher rates along with notably lower global disparity fractions compared with the ones mentioned above.

{
The time complexity of the decoding (encoding) algorithm of D-LOCO codes is algorithmically linear with respect to the code length $m$, i.e., $O(m)$. As we perform additions (subtractions) of at most $2m$-bit numbers at each iteration, implementation-wise we offer polynomial-time complexity, more precisely $O(m^2)$ (see Section~\ref{sec_balance} for encoding-decoding algorithms as well as Table~\ref{table3} for adder sizes).

From (\ref{storage_complexity}), the storage overhead is $O(m^2)$. Note that a lookup table technique to design a constrained code of length $m$ for DNA storage, where all valid sequences are stored, will have storage overhead of $O(4^m\cdot m)$, which is remarkably higher than $O(m^2)$.

\begin{remark} \label{offline_contributions} Note that symbol contributions can also be computed offline and used as inputs to the adder, provided that storage overhead requirements allow that. This results in an algorithmically linear decoder where only a single addition of at most $2m$-bit numbers is required at every iteration.
\end{remark}
}

\subsection{Reconfigurability} \label{reconfigurability}

Just like all LOCO codes \cite{ahh_general}, D-LOCO codes are also easily reconfigurable. In particular, the same set of adders used to encode-decode a specific D-LOCO code can be reconfigured to encode-decode another D-LOCO code of different length and/or run-length constraint. {Reconfigurability can play a pivotal role as the storage system ages and its performance gradually degrades. For instance, we may need to switch the run-length from $\ell$ to $\ell'$, where $\ell' < \ell$, in order to address the need of an aging, more error-prone system in which, homopolymers of shorter lengths become detrimental. While this will result in a rate loss, it can maintain the same level of reliability with time. Such reconfiguration is achieved simply by changing the inputs of the adders, i.e., the cardinalities, from $\{N_{\mathrm{D}}(i+1-\ell), \dots, N_{\mathrm{D}}(i)\}$ to $\{N_{\mathrm{D}}(i+1-\ell'), \dots, N_{\mathrm{D}}(i)\}$ through multiplexers to find the symbol contribution $g_i(c_i)$. Note that the system allows this reconfiguration since the required size of the adders gets lower as we move from $\ell$ to $\ell' < \ell$.

Provided that storage overhead requirements allow it, one can also compute the contributions of each symbol offline for $\mathcal{D}_{m, \ell}$ and $\mathcal{D}_{m, \ell'}$ and switch from the former to the latter through multiplexers in order to find the index of a codeword via $m$ additions (see Remark~\ref{offline_contributions}).
}
\subsection{Error Propagation and Parallelism} \label{error_propagation}

In D-LOCO codes, error propagation does not occur from a codeword into another. Moderate message lengths are preferred, however, in order to mitigate any possible codeword-to-message error propagation (see also \cite{ahh_loco} and \cite{ahh_general}). Furthermore, as they have fixed length, D-LOCO codes enable simultaneous encoding and decoding of different codewords within the complexity margin of the system, which can remarkably increase the speed of writing and reading. 

{
\subsection{Local Balancing} \label{local_balancing}

\begin{definition} (\hspace{-0.01em}\cite[Definition~2]{liu_he_tang}) A DNA sequence/stream of length $M$ satisfies \textit{$\mu$-prefix validity} if for every prefix $\mathcal{S}$ of it, i.e., for every subsequence consisting of the first $S$-many terms of it, $1 \leq S \leq M$, the disparity of $\mathcal{S}$ satisfies the inequality:
\begin{align} \label{prefix_validity}
\vert p(\mathcal{S}) \vert \leq 2\mu,
\end{align}
for non-negative $\mu$.

Recall that for a DNA sequence consisting of codewords in $\mathcal{D}_{m, \ell}$ of odd length $m$, when
\begin{itemize}
\item[-] Bridging Scheme~I and
\item[-] our balancing procedure 
\end{itemize}
are applied, its global disparity is in $[-m-1,m+1]$ (see Lemma~\ref{lemma_dispar}). Therefore, (\ref{prefix_validity}) is satisfied for all prefixes of length divisible by $m+1$ if we set $\mu = \lceil (m+1)/2 \rceil = (m+1)/2$. For the remaining lengths, we first write $\mathcal{S}$ as a concatenation of $\mathcal{S}_0$ and $\mathcal{S}_1$, where $\mathcal{S}_0$ is the subsequence up to the last bridging symbol, and $\mathcal{S}_1$ is the subsequence after the last bridging symbol. Without loss of generality, we consider the case of $\mathcal{S}_0$ having a positive disparity. Then, $\mu$ can be set to 
$$\lceil ((m+1)+(m-1)/2)/2 \rceil = \lceil (3m+1)/4 \rceil,$$
where the middle term $(m-1)/2$ is due to the maximum possible contribution of $\mathcal{S}_1$ to the disparity of $\mathcal{S}$ given that the disparity of $\mathcal{S}_0$ is at most $m+1$. Hence, such a DNA sequence satisfies $\lceil (3m+1)/4 \rceil$-prefix validity for $m \geq 5$. This remains true if we adopt Bridging Schemes~II-B or III instead (see Remark~\ref{rem_dispar}(ii) and Remark~\ref{rem_dispar}(iii)). 
\end{definition}

}

%%%%%%%%%%%%%%%%%%%%%%%%%%%%%%%%%%%%%
\section{Conclusion}\label{sec_conc}

We introduced D-LOCO codes, a family of constrained codes designed for DNA data storage. D-LOCO codes are equipped with lexicographic ordering of codewords, which enables simple bijective mapping-demapping between an index set and the codebook. We derived the mathematical rule governing this mapping-demapping, which leads to systematic, reconfigurable, and low-complexity encoding and decoding algorithms. We discussed four bridging schemes, three of which guarantee single substitution error detection, and also studied the probability of missing errors. We showed how D-LOCO codes can be balanced, locally and globally, with minimal rate loss. D-LOCO codes are capacity-achieving, and they have remarkably high finite-length rates at affordable complexities even with error-detection and balancing features. Future work includes extending D-LOCO codes to forbid other detrimental patterns as well as combining constrained codes with error-correction codes for DNA data storage.

%%%%%%%%%%%%%%%%%%%%%%%%%%%%%%%%%%%%%
\section*{Appendix} \label{appendix}

In this appendix, we present a \textit{generalized} version of D-LOCO codes, namely GD-LOCO codes, for any alphabet size.

\begin{definition} For an integer $q \geq 2$, let $E=\{\Delta^r \textrm{ } | \textrm{ } 0 \leq r \leq q-1\}$ be a lexicographically-ordered alphabet with $q$ elements such that $\Delta^r < \Delta^s$ for $0\leq r < s \leq q-1$. For $\ell \geq 1$, the GD-LOCO code $\mathcal{GD}_{m,\ell}$ is defined as the set of all codewords of length $m$ defined over $E$ that do not contain runs of length exceeding $\ell$. 

We denote the cardinality of $\mathcal{GD}_{m,\ell}$ by $N(m)$ (instead of $N_{\mathrm{GD}}(m,\ell)$) for the ease of notation.
\end{definition} 

\begin{proposition} \label{prop_card_general} (\hspace{-0.01em}\cite[Equation~1]{immink_cai}) The cardinality $N(m)$ of the GD-LOCO code $\mathcal{GD}_{m,\ell}$, where $\ell \geq 1$, satisfies the following recursive relation for $m \geq \ell$: 
\begin{align} \label{card} N(m) &= (q-1) \sum^{\ell}_{i=1} N(m-i).
\end{align}
For $0 \leq m \leq \ell$, $$N(0) \triangleq \frac{q}{q-1}, \textrm{ and } N(m)=q^m \textrm{ for } 1 \leq m \leq \ell.$$
\end{proposition}

\begin{theorem} \label{thm_mostgeneral} The encoding-decoding rule $g: \mathcal{GD}_{m,\ell} \rightarrow \{0,1,\dots,N(m)-1\}$, for general $\ell \geq 1$, is as follows:
\begin{equation}
g(\mathbf{c})= \frac{q-1}{q} \sum^{m-1}_{i=0} \sum^{\ell}_{j=1} \sum^{j}_{k=1} \bigg( \sum_{\Delta \in E^*} \delta_{i,k} \bigg) N(i+j-\ell), 
\end{equation}
\\
where $N(0) \triangleq q/(q-1)$, $N(-1) \triangleq \dots \triangleq N(2-\ell) \triangleq N(1-\ell) \triangleq 0$, and $E^* = E \setminus \{\Delta^{q-1}\}$. Moreover, for all $\Pi > \Delta$ and for each $\Delta \in E \setminus \{\Delta^{q-1}\}$,
\begin{align} \delta_{i,1} &=1 \textrm{ only if } c_i=\Pi \textrm{ and } c_{i+1} \neq \Delta, \nonumber \\
\delta_{i,k} &=1 \textrm{ only if } c_{i+k-1}\dots c_{i+1}c_i=\mathbf{\Delta}^{k-1} \Pi \textrm{ and } c_{i+k} \neq \Delta,
\end{align}
where $2 \leq k \leq \ell$ (these are $\ell-1$ statements), and in all other cases, $\delta_{i,k}=0$ for $i \in \{1,2,\dots,m\}$ and $k \in \{1,2,\dots,\ell\}$. \\
Here, $\Pi = \Delta^s > \Delta = \Delta^r$, for some $0 \leq r < s \leq q-1$, is according to the lexicographic ordering rule, and $\delta = \delta^r$ in $\{\delta^0, \delta^1, \dots, \delta^{q-2}\}$ stands for the small letter corresponding to $\Delta = \Delta^r$ in $\{\Delta^0, \Delta^1, \dots, \Delta^{q-2}\}$.

{Recall that $c_i \triangleq \zeta$ (a letter outside of the alphabet), for all $i > m-1$ (see Remark~\ref{out_of_codeword}).}
\end{theorem}

\begin{remark} \label{remark_RLL} For $q=2$, the alphabet $E=\{\Delta^0,\Delta^1\}$ can be defined as the Galois field $\mathrm{GF}(2) = \{0,1\}$. Lexicographically-ordered RLL (LO-RLL) codes can be designed as shown in \cite{tang_bahl}. For $\ell \geq 2$, define the difference vector $\mathbf{v}$ of a GD-LOCO codeword $\mathbf{c}$ in $\mathcal{GD}_{m,\ell}$ as $\mathbf{v}=[v_{m-2}\,v_{m-3}\,\dots\,v_0]$, with $v_i=c_{i+1}+c_i$ over $\mathrm{GF}(2)$, for all $i \in \{0,1,\dots,m-2\}$. Observe that for any codeword $\mathbf{c}$ in $\mathcal{GD}_{m,\ell}$, its difference vector satisfies the $(0,\ell-1)$ RLL constraint. Moreover, each codeword in the $(0,\ell-1)$ LO-RLL code of length $m-1$ can be derived from a unique codeword among the GD-LOCO codewords starting with $0$ from the left in $\mathcal{GD}_{m,\ell}$ by computing the difference vectors of all such codewords (the difference vectors of the remaining codewords will be duplicates). Consequently, the cardinality $N_{\mathrm{RLL}}(m-1,\ell-1)$ of the $(0,\ell-1)$ LO-RLL code is given by:
\begin{equation}
N_{\mathrm{RLL}}(m-1,\ell-1)=\frac{1}{2}N_{\mathrm{GD}}(m,\ell).
\end{equation}
This observation leads to a simple way of constructing and indexing $(0,\ell-1)$ LO-RLL codes via GD-LOCO codes with parameters $q=2$ and $\ell$.
\end{remark}
%%%%%%%%%%%%%%%%%%%%%%%%%%%%%%%%%%%%%

\begin{remark} For $q=16$, consider the GD-LOCO code $\mathcal{GD}_{m,1}$ consisting of codewords with no identical consecutive symbols. Note that if we convert codewords in $\mathcal{GD}_{m,1}$ to $4$-ary codewords, where each $16$-ary symbol is mapped to a unique $2$-tuple of $4$-ary symbols, we obtain a set of $4$-ary codewords of length $2m$ that do not contain any patterns of the form $\Lambda_1\Lambda_2\Lambda_1\Lambda_2\Lambda_1$, where $\Lambda_1$ and $\Lambda_2$ can be the same symbol in the alphabet of size $4$. If this alphabet is $\{A,T,G,C\}$, such $4$-ary coding scheme allows runs of length at most $4$ and also eliminates short tandem repeats such as $AGAGAG$.
\end{remark}

\section*{Acknowledgment}
The authors would like to thank \"{O}zge Simay Demirci and Selin S\"{o}nmez for their assistance in carrying out this research. Furthermore, they would like to thank the Guest Editor Prof. Eitan Yaakobi for his effective handling of the article, and thank the anonymous reviewers for their constructive comments.

%%%%%%%%%%%%%%%%%%%%%%%%%%%%%%%%%%%%%

%%%%%%%%%%%%%%%%%%%%%%%%%%%%%%
\balance

\begin{IEEEbiographynophoto}{Canberk \.{I}rima\u{g}z{\i}} received the B.Sc. degree in Mathematics from Ko\c{c} University, Turkey, and the M.A. degree in Mathematics from the University of Wisconsin-Madison, USA. He is currently pursuing the Ph.D. degree in Cryptography at the Institute of Applied Mathematics, Middle East Technical University (METU), Turkey. His current research interests include applied algebra, coding theory, and DNA data storage.
\end{IEEEbiographynophoto}

\begin{IEEEbiographynophoto}{Yusuf Uslan} is an undergraduate student at the Electrical and Electronics Engineering Department of Middle East Technical University (METU), Turkey. His research interests include coding theory, signal processing, and DNA data storage.
\end{IEEEbiographynophoto}

\begin{IEEEbiographynophoto}{Ahmed Hareedy} (Member, IEEE) is an Assistant Professor with the Department of Electrical and Electronics Engineering at Middle East Technical University (METU), Turkey. He is also an Affiliated Faculty Member with the Institute of Applied Mathematics at METU, Turkey. He is interested in questions in coding/information theory that are fundamental to opportunities created by the current, unparalleled access to data and computing. He received the Bachelor and M.S. degrees in Electronics and Communications Engineering from Cairo University, Egypt, in 2006 and 2011, respectively. He received the Ph.D. degree in Electrical and Computer Engineering from the University of California, Los Angeles (UCLA), USA, in 2018. He was a Postdoctoral Associate with the Department of Electrical and Computer Engineering at Duke University, USA, between 2018 and 2021. He worked with Mentor Graphics Corporation (currently, Siemens EDA) between 2006 and 2014. He worked as an Error-Correction Coding Architect with Intel Corporation in the summers of 2015 and 2017.

Dr. Hareedy won the 2018--2019 Distinguished Ph.D. Dissertation Award in Signals and Systems from the Department of Electrical and Computer Engineering at UCLA. He is a recipient of the Best Paper Award from the 2015 IEEE Global Communications Conference (GLOBECOM), Selected Areas in Communications, Data Storage Track. He won the 2017--2018 Dissertation Year Fellowship (DYF) at UCLA. He won the 2016--2017 Electrical Engineering Henry Samueli Excellence in Teaching Award for teaching Probability and Statistics at UCLA. He is a recipient of the Memorable Paper Award from the 2018 Non-Volatile Memories Workshop (NVMW) in the area of devices, coding, and information theory. He is a recipient of the 2018--2019 Best Student Paper Award from the IEEE Data Storage Technical Committee (DSTC). He has been awarded the T\"{U}B\.{I}TAK 2232-B International Fellowship for Early Stage Researchers in 2022. He is currently a Guest Editor of the Special Issue on Data Storage of the {\sc IEEE BITS the Information Theory Magazine}.
\end{IEEEbiographynophoto}

\end{document}